\documentclass[reprint,amsmath,amssymb,txfonts,aps,pre,superscriptaddress,showpacs,floatfix]{revtex4-1} 
\usepackage{graphicx,bm,amssymb,amsmath,dcolumn,hyperref}
\usepackage{color,multirow}
\usepackage{braket}
\usepackage{mathrsfs}

\usepackage{float}
\usepackage{ulem}
\begin{document}
\definecolor{red}{rgb}{1,0,0}
\newcommand{\red}[1]{\textcolor{red}{#1}}
\newcommand{\muast}{\mu^\ast}
\preprint{APS/123-QED}

\title{Charge density waves and superconductivity in the electron-positive fermion gas using a simple intuitive model. Part II: Collective modes, effective interactions, superconductivity, and transport.}


\author{Carl A. Kukkonen} 
\email{kukkonen@cox.net}
\affiliation{33841 Mercator Isle, Dana Point California 92629, USA}


\date{\today}

\begin{abstract}

Superconductivity and the normal state electrical resistivity which varies as $T^2$ are strongly enhanced near the compressibility and charge density wave instabilities in the electron-positive fermion gas. The additional screening from the positive fermions introduces an attractive term in the effective electron-electron interaction that is the basis for superconductivity. Electron-positive fermion scattering is the source of the $T^2$ term in the electrical resistivity. At an instability, both interactions are divergent. 

The previous paper Part I \cite{ref1} derives the formulas for the interactions and examines the source of the charge density waves and phase diagram of the simple model used for the electron-positive fermion gas. This paper uses local field factors consistent with Quantum Monte Carlo calculations in these formulas to calculate numerical results for the effective interactions amongst the electrons and positive fermions. The collective modes are also calculated and found to terminate (soften) due to the charge density waves. 

The superconducting transition temperature is estimated using the McMillan formula. The electron-positive fermion gas conducts electricity and heat. Because electron-electron and positive fermion-positive fermion scattering conserve momentum, they do not contribute to the electrical resistivity, but electron-positive fermion scattering does. All three scattering mechanisms contribute to the thermal resistivity.

The simple model for the electron-positive fermion gas is physically intuitive and naturally introduces instabilities at $q=0$ when the bulk modulus becomes zero and charge density waves at finite $q$ under some circumstances. For each mass ratio $M/m$, there is a unique density $r_s$ where the energy is a minimum. For different mass ratios, the interactions are investigated at several values of $r_s$ ranging from below the energy minimum to that of the instability.

How to potentially map the simple model onto real materials, the nature of the new phases, and several missing elements and contradictions are beyond the talents of the author, and invite future work.

\end{abstract}

\maketitle
\section{Introduction}

This paper is a direct follow-on to the previous paper by the same author \cite{ref1} that used a simple intuitive model for the electron-positive fermion gas together with current local field factors to predict that $q=0$ instability and charge density waves are inherent in the two component gas. The charge density waves appear because of a combination of the additional screening by the positive fermions and the use of local field factors derived from Quantum Monte Carlo calculations. These local field factors continue to rise as $q^2$ up to almost $q = 2_{kF}$ which leads to a divergence in the response function at finite $q$, as discussed in \cite{ref1}.  

I use the simple model with these local field factors to calculate the collective modes, the electron-positive fermion effective interaction and the effective interactions between two electrons and between two positive fermions. The results show the impact of the compressibility instability at $q=0$ and the charge density waves. The additional screening from the positive fermions yields an attractive effective interaction term between two electrons and leads to BCS like superconductivity.  This electron-positive fermion interaction term leads to a $T^2$ temperature dependence in the normal state electrical resistivity. Both diverge at the instabilities and can lead to predictions of high temperature superconductivity as well as large normal state resistivity.

The basic physics concepts were understood in the $1970$s. Vashishta, Bhattacharyya and Singwi \cite{ref2} presented the linear response theory of the electron-hole liquid. Kukkonen and Overhauser \cite{ref3} calculated the electron-electron interaction in a deformable background and expressed the result in terms of then unknown local field factors. For quantitative calculations, most previous authors used the RPA, Thomas Fermi and more sophisticated approximations for the needed interactions. Vignale and Singwi \cite{ref4,ref5,ref6,ref7} considered collective modes and superconductivity using the Hubbard approximation which has the correct behavior at $q=0$, but not at finite $q$, and does not predict charge density waves. Richardson and Ashcroft \cite{ref8} also investigated superconductivity in the electron-hole gas and did an independent calculation of the local field factors. These calculations were difficult and they did not find charge density waves. Kukkonen and Maldague \cite{ref9,ref10,ref11,ref12} derived the transport properties of the electron-hole gas in terms of the scattering rate of electrons by holes. 

The best available local field factors \cite{ref13} satisfy the compressibility and susceptibility sum rules at small wave vector and follow the results found by Quantum Monte Carlo calculations at intermediate wave vector, and satisfy another sum rule at large wave vector. 

The simple model assumes that the positive fermions form a uniform background for the electrons and vice versa. Thus the considerable body of work on the electron gas in a uniform background can be directly applied. In calculating the total energy of the electron-positive fermion gas, it is assumed that the total energy is simply the sum of the energy of the electron gas uniform background and a positive fermion gas in a uniform background. Exchange and correlation within each species is accounted for, but the electron-positive fermion correlation energy is not included. I argued in the previous paper \cite{ref1} that this energy was likely nearly a constant as a function of density and therefore does not contribute to the response functions. Calculation of this correlation energy is beyond my capabilities, and remains an open issue that could validate or disprove this assumption of the simple model. 

The collective modes called acoustic plasmons are calculated in Section II, and compared to previous work. The charge density waves terminate the dispersion relation of the acoustic plasmons.  The electron-positive fermion interaction is addressed in Section III. Electron-electron and positive fermion-positive fermion scattering do not contribute to the electrical resistivity because their contribution to the current is proportional to their total momentum that is conserved in scattering. The total current in electron-positive fermion gas is not proportional to momentum and is degraded by electron-positive fermion scattering which transfers momentum from one species to the other. The electron-positive fermion interaction also plays the role of the electron-phonon interaction in BCS superconductivity.

 Section IV addresses the effective electron-electron interaction and the positive fermion-positive fermion interaction is considered in Section V. Previous results are re-derived and the current local field factors are used to produce quantitative results. The effective electron-electron interaction is the sum of two terms. The first term is the effective electron-electron interaction in a uniform background. The second attractive term is proportional to the electron-positive fermion interaction which becomes very large in the density region close to compressibility and charge density wave instabilities.

The electrons and positive fermions have no magnetic/spin coupling and the induced magnetization is just the sum of the magnetization of the fermions considered separately as discussed in Section VI. Superconductivity is considered in Section VII.  In BCS superconductivity, the repulsive portion of the effective electron-electron interaction inhibits superconductivity and is often represented by the superconducting parameter $\mu$. The attractive portion of the interaction provided by the effective electron-positive fermion interaction is used to calculate the attractive superconducting parameter $\lambda$. At densities near an instability, $\lambda$ becomes large. The McMillan formula is used to estimate the superconducting transition temperature. 

The transport properties of the normal state are discussed in Section VIII. The electrical resistivity of the normal state is proportional to the scattering rate of electrons by positive fermions. This is an angular average of the effective electron-positive fermion interaction squared. The resistivity is proportional to $T^2$ and can become very large if the density is close to an instability.

The overall results of this investigation of the electron-positive fermion gas are summarized in Section IX and some suggestions for future work to validate, disprove or improve the simple model are made. The limitations of the simple model and speculations about potential applicability are discussed in Appendix A.

\section{Collective modes}

Vignale and Singwi \cite{ref4,ref5,ref6} addressed the collective modes of electron-hole liquids using the Hubbard approximation, which was the best available local field factor at the time. They worked in the center of mass frame using the reduced mass and effective $r_s$.  In this section, I compare the collective modes using the intuitive simple model using the actual density, and current best local field factor to the RPA and Hubbard approximation for a wide range of mass ratios. 

The expression for the dispersion relation $\omega_q$ of the collective modes was given in Eq. (C10) of paper I \cite{ref1} and is repeated below
\begin{eqnarray}\label{eq1}
    \omega_q^2  =   \frac{\omega_{p1}^2}{v} \frac{V_{11}^b(1 + G_{1+})+(V_{12}^b(1 - 2G_{12}))^2\Pi^0_2}{\varepsilon_{\text{et}}}       
\end{eqnarray}

The subscript 1 refers to the positive fermions and 2 to the electrons. $G$ is the local field factor, $v = 4\pi e^2/q^2$ is the coulomb interaction,   $\Pi^0_2$ is the Lindhard function for electrons and $\varepsilon_{\text{et}}$ is the electron test charge dielectric function. The plasma frequency for the positive fermions $\omega_{p1}^2$ can be written in several different equivalent ways:
\begin{eqnarray}\label{eq2}
    \omega_{p1}^2 &=& = \frac{4 \pi e^2 n}{M} = \frac{ mv_{F2}^2 q_{TF}^2 }{3M} = c_{BS}^2 q_{TF}^2 \; \; ,  
\end{eqnarray}
where $n$ is the electron and positive fermion density, $m$ is the electron mass, $M$ is the positive fermion mass, $v_{F2}$ is the electron Fermi velocity, $q_{TF}$ is the Thomas-Fermi wave vector of the electrons, and $c_{BS}$ is the Bohm-Staver speed of sound \cite{ref14}. 

The superscript $b$ refers to the bare interaction which allows for a pseudopotential. Specializing to the simple case where all interactions are coulomb interactions $v$, and ignoring the electron-positive fermion correlation by setting $G_{12}=0$, Eq. \eqref{eq1} becomes
\begin{eqnarray}\label{eq3}
\omega_q^2 =   \omega_{p1}^2 \frac{(1- G_{1+}) - v \Pi^{0\text{e}}}{\varepsilon_{\text{et}}} \; . 
\end{eqnarray}
Using the equality for the test charge-test charge dielectric function in the electron gas, $1- 1/\varepsilon_{\text{tt}} = v \Pi^{0\text{e}}/ \varepsilon_{\text{et}}$, the dispersion relation can be rewritten as
\begin{eqnarray}\label{eq4}
\omega_q^2 =  \omega_{p1}^2 \left( \frac{1}{\varepsilon_{\text{tt}}} - G_{1+} \right) \; .          
\end{eqnarray}

The first term in this equation is simply the dispersion relation in the electron gas which includes exchange and correlation of the electrons, but simply treats the positive fermions as classical charged point masses. The second term reflects that the positive fermions are indeed fermions and have exchange and correlation interactions with each other.

In the random phase approximation RPA, all of the local field factors equal zero and the simple textbook definition of the dispersion relation is obtained \cite{ref13}. At small $q$ the RPA dispersion relation is $\omega_q = (m /3M)^{1/2} v_{F2} q  = c_{BS} q$ and the speed of sound is simply the Bohm-Staver value. 

Including exchange and correlation, the speed of sound is reduced substantially below the Bohm-Staver value and becomes imaginary for $r_s$ greater than approximately $2.5$. The precise value varies slightly with the mass ratio. This result depends only on the compressibility sum rule and was previously obtained in Ref. \cite{ref4}. For every mass ratio $M/m$, the electron-positive fermion gas has a unique energy minimum. For mass ratios $M/m < 3.5$, the energy minimum is at $r_s > 2.5$ (see Ref. \cite{ref15}), and the collective mode does not exist. In Ref. \cite{ref4}, the authors noted that they found no solution at the energy minimum density for a collective mode for $M/m = 3$ which is consistent with the observation above. 

\begin{figure}
    \centering
    \includegraphics[width=1\linewidth]{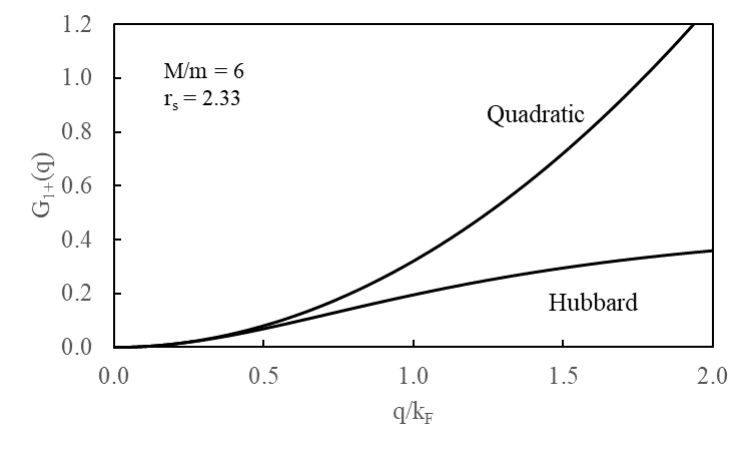}
    \caption{Density local field factor $G_{1+}(q)$ of the positive fermions plotted as a function of wave vector at a representative mass ratio and the $r_s$ of the unique energy minimum at that mass ratio. Quantum Monte Carlo calculations show that $G_{1+}(q)$ continues as approximately $q^2$ up to $2 k_F$, where the Hubbard approximation approaches a constant of $1/2$.}
    \label{fig1}
\end{figure}

The collective mode is very sensitive to the local field factor at finite wave vector. Any local field factor that satisfies the compressibility sum rule will give the same speed of sound at small $q$.  Reference \cite{ref4} used the Hubbard approximation where $G_+ = q^2/(2(q^2 +C))$ where the constant is chosen to satisfy the compressibility sum rule and goes to a constant 1/2 at large $q$. The Hubbard approximation begins to fall significantly below $q^2$ for $q> 0.7 \, k_F$. Quantum Monte Carlo calculations have shown that $G_+$ remains closely proportional to $q^2$ up to nearly $q=2 \, k_F$. The Hubbard approximation is compared to the quadratic Quantum Monte Carlo result in Fig.\ref{fig1}. 

Including exchange and correlation in the electron gas, reduces $1/\varepsilon_{\text{tt}}$ slightly, and exchange and correlation in the positive fermion gas means that $G_{1+}$ increases as $q^2$. The result is that at a certain wave vector $\omega_q^2$ in Eq. \eqref{eq4} becomes negative and the collective mode ceases to exist. This $q^2$ behavior is what causes the charge density waves as well.

\begin{figure}
    \centering
    \includegraphics[width=1\linewidth]{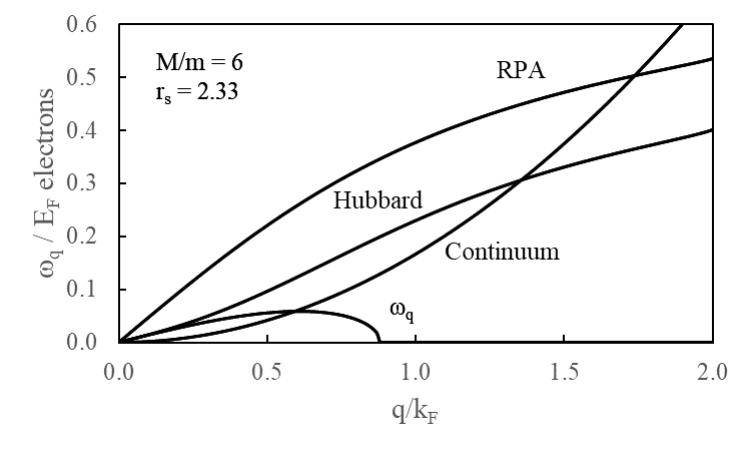}
    \caption{Illustrative example of the collective mode frequency divided by the electron Fermi energy as a function of wave vector. The Random Phase Approximation (RPA) and Hubbard approximation are compared to the collective mode frequency $\omega_q$ calculated using the correct local field factor that follows the $q^2$ behavior determined by Quantum Monte Carlo calculations. The edge of the particle-hole continuum of the heavier positive fermions, $\hbar^2 q^2/2M$, is also shown.}
    \label{fig2}
\end{figure}

 Using the simple model and the known local field factors, it is trivial to plot $\omega_q$ for different mass ratios $M/m$ and densities $r_s$. An illustrative example is given in Fig. \ref{fig2}.


The difference between collective mode dispersion relation $\omega_q$ calculated using the Quantum Monte Carlo local field factor and that of the Hubbard approximation in Fig. \ref{fig2} is striking. The Hubbard local field factor never gets large enough to cancel the first term in Eq. \eqref{eq4}. This behavior might be experimentally observable and provide verification of the intermediate wave vector behavior of the local field function.

The collective mode is well-defined until the frequency crosses the particle-hole continuum. At that point, damping begins and the collective mode loses clear definition. Figure \ref{fig1} illustrates several important points. First, the RPA predicts the largest collective mode with linear behavior at small $q$ with the slope being the well-known Bohm-Staver speed of sound (see Eq. \eqref{eq2}. The Hubbard approximation includes a local field factor that satisfies the compressibility sum rule at $q=0$ and that drastically reduces the slope at small $q$ (which is the speed of sound). However the Hubbard approximation to the local field factor does not have the $q^2$ behavior at intermediate and large $q$, and this leads to a serious overestimate of the collective mode in addition to not predicting charge density waves.

The collective mode predicted by using the current local field factor is shown by the curve $\omega_q$ in Fig.\ref{fig1}. Although, $\omega_q$ starts with the same slope as the Hubbard approximation, it quickly falls away from linearity in wave vector and crosses the particle-hole continuum at a much lower wave vector. There is a definite region of a stable collective mode, but it is much smaller than predicted by the Hubbard approximation, and much much smaller than the RPA prediction. 

The same trend is true for acoustic plasmons in a system with two different mass electrons. When Pines \cite{ref15} first considered this problem in 1956, he suggested that acoustic plasmons be named demons in honor of Maxwell. The fact that the acoustic plasmon mode is much much smaller than predicted by Pines using the RPA may the reason that they have been so hard to see experimentally \cite{ref16}.

\begin{figure}
    \centering
    \includegraphics[width=1\linewidth]{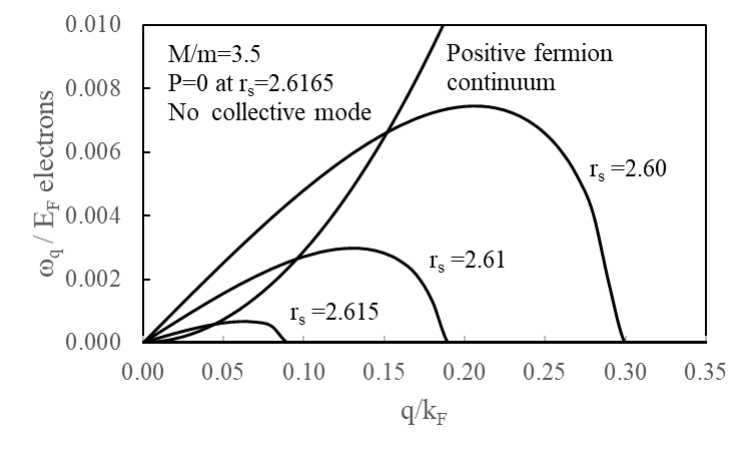}
    \caption{Collective mode frequency at $M/m=3.5$ divided by the Fermi energy of the electrons as a function of wave vector. $M/m=3.5$ is approximately the lowest mass ratio where the collective mode occurs at the $r_s$ of the energy minimum. At $r_s= 2.6165$, the collective mode is infinitesimal. Increasing the density (reducing $r_s$) sharply increases the collective mode. Also shown is the particle-hole continuum of the heavier positive fermions.}
    \label{fig3}
\end{figure}

Near the minimum mass ratio $M/m=3.5$ for a collective mode at the $r_s$ corresponding to the energy minimum of the electron-positive fermion gas, the collective mode dispersion relation $\omega_q$ is extremely sensitive to $r_s$. At the energy minimum $r_s = 2.6165$, there is no collective mode. However a well-defined collective mode develops quickly as density increases and is well developed at $r_s =2.61$ as shown in Fig. \ref{fig3}. The slope of the dispersion relation (sound speed) varies quickly with $r_s$.  The pressure dependence of acoustic plasmons should be interesting if observable. 

\begin{figure}
    \centering
    \includegraphics[width=1\linewidth]{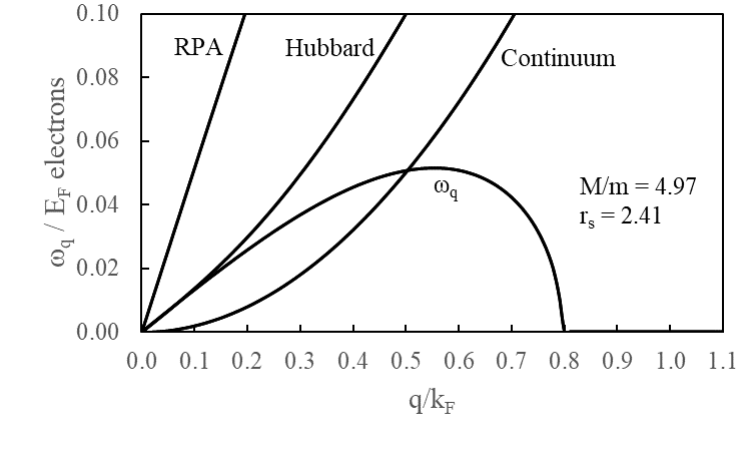} \includegraphics[width=1\linewidth]{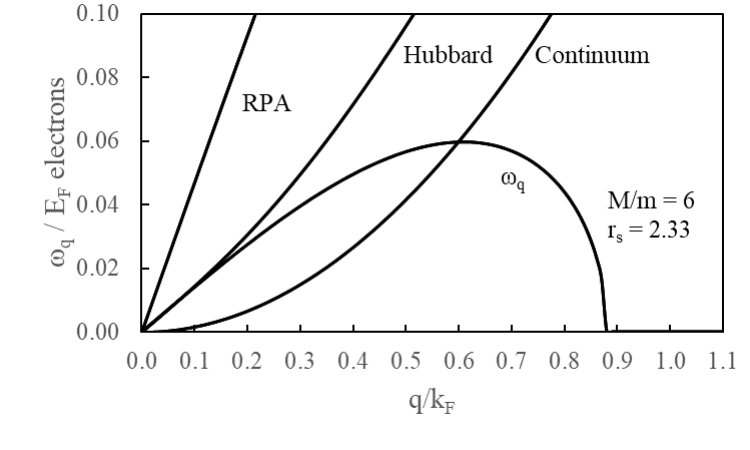}
    \includegraphics[width=1\linewidth]{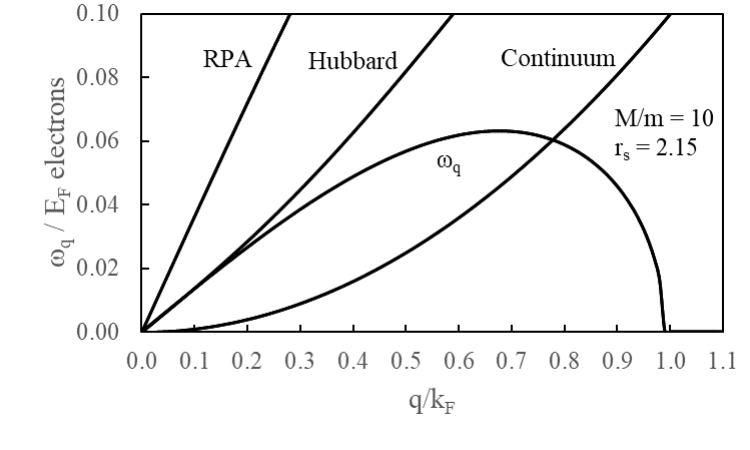} \includegraphics[width=1\linewidth]{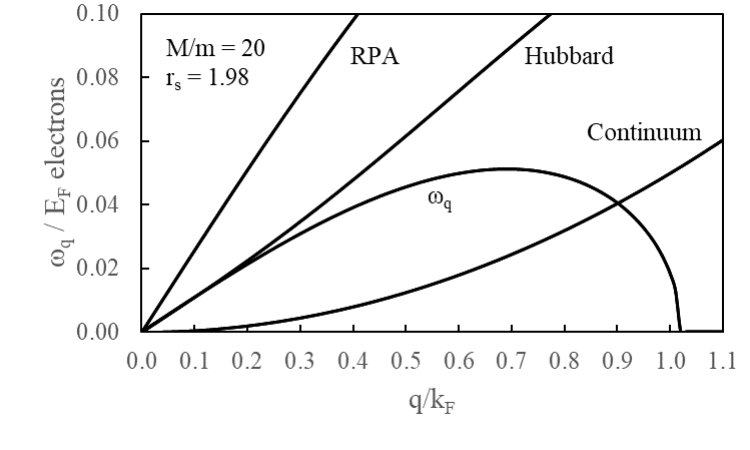}
    \caption{Predicted collective mode $\omega_q$ divided by the Fermi energy of the electrons plotted versus wave vector for representative values of the mass ratio $M/m$. The collective modes predicted by the RPA and Hubbard approximation are shown along with the boundary of the particle-hole continuum. $\hbar^2 q^2/2M$, of the heavier positive fermions. Each plot is made at the equilibrium $r_s$ where the energy of the electron-positive fermion gas is a minimum.}
    \label{fig4}
\end{figure}

The calculation of the equilibrium density \cite{ref1} was done assuming that both the electrons and positive fermions were in a uniform background. In the simple model, the total energy is simply the total of the electron gas in a uniform background and the positive fermion gas in a uniform background, and any additional correlation energy is ignored. The energy is well known and the unique minimum is be found by varying $rs$. When calculating the response functions, charge density waves at finite wave vector are found to begin at certain values of $r_s$ that differ from the density of the energy minimum (see Fig. 15 of Ref. \cite{ref1}). There are no charge density waves for mass ratios $M/m<4.97$. For $4.97< M/m <9$, charge density waves are predicted at densities lower than the equilibrium density, but higher than the density of the $q=0$ instability. For mass ratios $M/m > 9$, charge density waves are predicted at the equilibrium density.

The simple model predicts collective modes at the equilibrium density for all mass ratios $M/m>3.5$. The collective modes all terminate at wave vectors below the wave vector of the charge density wave. I see no direct influence of the charge density wave on the collective mode. This may be obvious, but not to me. Although with a charge density wave, the background is no longer uniform and the model assumption is violated, Vignale and Singwi \cite{ref6,ref7} argue that the presence of the charge density wave does not prevent superconductivity. They also find that collective modes and superconductivity are compatible.

Figure \ref{fig4} shows the collective modes calculated for representative mass ratios between $4.97$ and $20$ at their respective equilibrium densities.

There is a well-defined collective mode at the energy minimum $r_s$ for $M/m > 3.5$. These collective modes cross the edge of the particle-hole continuum of the positive fermions at wave vectors $q/k_F < 0.95$, and completely terminate at wave vectors $q/k_F< 1.1$. 

\section{ELECTRON-POSITIVE FERMION INTERACTION}
The general equations defining the effective interactions were derived in Appendix C of Paper I \cite{ref1}. They are given below.
\begin{eqnarray}
V_{1+}^{\text{eff}} &=& \left(V_{1+}^{\text{ext}} + V_{12}^b (1- 2G_{12}) \Delta n_{2+}\right)\frac{1}{\varepsilon_{\text{ht}}} \label{eq5}\\
V_{2+}^{\text{eff}} &=& \left( V_{2+}^{\text{ext}} + V_{21}^b (1- 2G_{21}) \Delta n_{1+} \right)\frac{1}{\varepsilon_{\text{et}}} \label{eq6}
\end{eqnarray}

The response of spin up and spin down fermions are different when the external potential is spin dependent. As shown in Paper I, it is natural to add and subtract the responses of spin up and spin down fermions, because the density response only depends on the sum, and the magnetic response on the difference of the induced densities. The subscript $1+$ indicates the sum of the external and effective potentials of positive fermions with spin up and spin down. The subscript 2 refers to the electrons. $V_{12}^b$ is the bare interaction between the electrons and positive fermions. In general, it can be a pseudopotential, but for the calculations below, it taken as simply the coulomb interaction. $G_{12} = G_{21}$ is the unknown electron-positive fermion correlation local field factor. 

Linear response from the electron gas yields the relationship between the induced density and the effective potential, $\Delta n_{2+} = -\Pi^0_2 V_{2+}^{\text{eff}}$ and $\varepsilon_{\text{et}} = 1 + v(1-G_{2+}) \Pi^0_2$, with analogous  definitions for the positive fermions with dielectric function $\varepsilon_{\text{ht}}$. The Fourier transform of the coulomb interaction is $v = 4 \pi e^2/q^2$.

The key to using Eqs. (\ref{eq5},\ref{eq6}) is to determine the external potential that is perturbing the system. To calculate the electron-positive fermion effective interaction, the perturbation is taken to be a positive fermion with spin up. Since this positive fermion is identical to other positive fermions with spin up, there is an exchange and correlation interaction with them. There is a correlation interaction with spin down positive fermions, and there is a correlation interaction with electrons. This is the distinction between the disturbance being a simple test charge at a fixed location (effectively with infinite mass) which is distinguishable from all of the particles in the gas and does not move (no correlation). The positive fermion has finite mass and has a correlation interaction with both species.

The external disturbance of a positive fermion with spin up and written as $\rho_{1\uparrow}$. See the discussion of $G_{\uparrow \uparrow}$ the spin dependent local field factors in Appendix C of Ref. \cite{ref1}. The external potential felt by another spin-up positive fermion is $V_{1\uparrow}^{\text{ext}} = v \rho_{1\uparrow}(1 - 2G_{1\uparrow \uparrow})$. The external potential felt by a positive fermion with spin down is $V_{1\downarrow}^{\text{ext}} = v \rho_{1\uparrow}(1- 2G_{1 \downarrow \uparrow})$ and $V_{1+}^{\text{ext}} = v \rho_{1\uparrow}(1- G_{1 +})$. Similarly the external potential by the electrons with spin up and spin down are equal and $V_{2+}^{\text{ext}} = -v \rho_{1\uparrow }(1-2G_{12})$, assuming $G_{12}= G_{21}$ but attractive rather than repulsive. With these external potentials and the linear response relationship between the effective potential and the induced density, Eqs. (\ref{eq5}-\ref{eq6}) become two equations with two unknowns, and are solved to yield
\begin{eqnarray}\label{eq7}
V_{21}^{\text{eff}} &=& V_{2+}^{\text{eff}} = - v \rho_{1\uparrow} \frac{1- 2G_{12}}{\varepsilon_{\text{et}} \varepsilon_{\text{ht}} – v^2 \Pi^0_1 \Pi^0_2 (1-2G_{12})^2} \nonumber  \\
&=& - \frac{v \rho_{1\uparrow}(1- 2G_{12}) }{\Delta}  \; \; . 
\end{eqnarray}

The effective potential felt by positive fermion due to an electron disturbance is the same because $\Delta$ is symmetric with respect to interchange of electrons and positive fermions. The quantity $\Delta * (q/k_F)^2$ was extensively examined in Paper I \cite{ref1} as a function of mass ratio $M/m$ and density $r_s$. For every mass ratio, $\Delta * (q/k_F)^2= 0$ at $q=0$ for a certain $r_s$ which signifies the compressibility instability of the electron-positive fermion gas. For mass ratios $M/m>4.97$, there is an additional zero at finite $q/k_F$ at a different $r_s$ which signifies a charge density wave. The zeros of $\Delta * (q/k_F)^2$ are divergences in the effective electron-positive fermion interaction.

For each mass ratio, the simple model predicts a unique $r_s$ where the energy of electron-positive fermion gas is a minimum and the gas is in equilibrium. Compression would be needed to get to a smaller $r_s$ and a pseudopotential to get to a larger $r_s$. The electron-positive fermion gas has two parameters $r_s$ and the mass ratio $M/m$. Since this simple model gives formulas for all of the interactions, it is easy to investigate the dependence on these parameters.

In the past, various approximations have been made for the electron-positive fermion interaction.
One option that ignores the additional screening by the positive fermions is the electron-test charge interaction in the electron gas with a uniform background $v/\varepsilon_{\text{et}}$. This is equal to the Thomas-Fermi interaction at $q=0$. Another option is to include both electron and positive fermion screening in the random phase approximation $v/\varepsilon_{RPA}$, where $\varepsilon_{RPA} = 1 + v \Pi^0_1 + v \Pi^0_2$. A third option is to use the Hubbard approximation which satisfies the compressibility sum rule at $q=0$, but does not agree with the quantum Monte Carlo results at larger $q$. These three alternatives are compared at three different densities in Fig. \ref{fig5} with $V_{21}^{\text{eff}}$, the result using the simple model and the current local field factors. 

\begin{figure}[h!]
    \centering
    \includegraphics[width=1\linewidth]{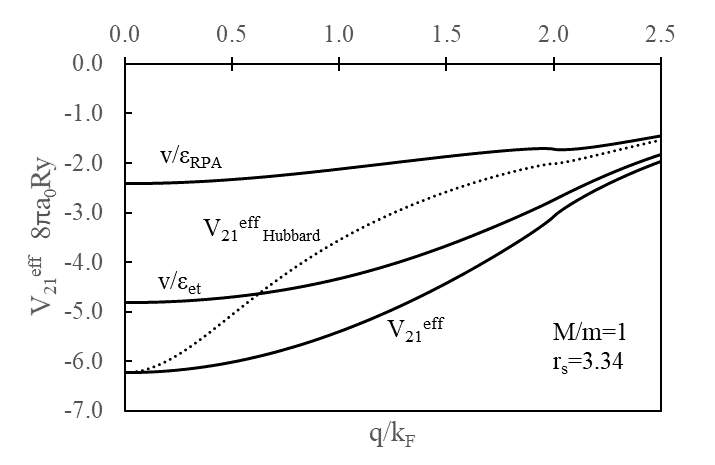}
    \includegraphics[width=1\linewidth]{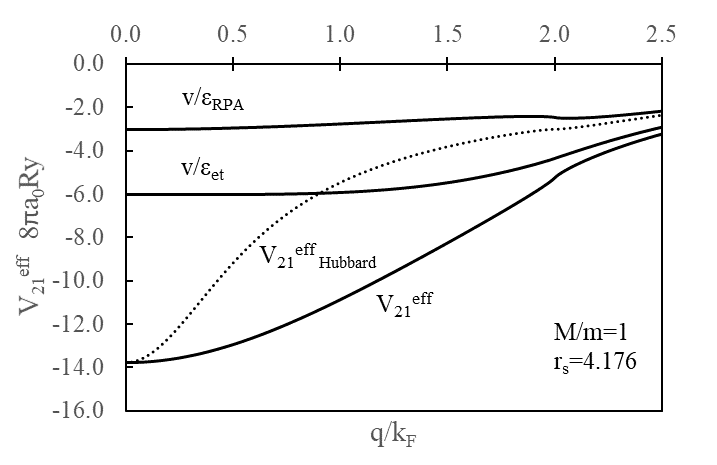}
    \includegraphics[width=1\linewidth]{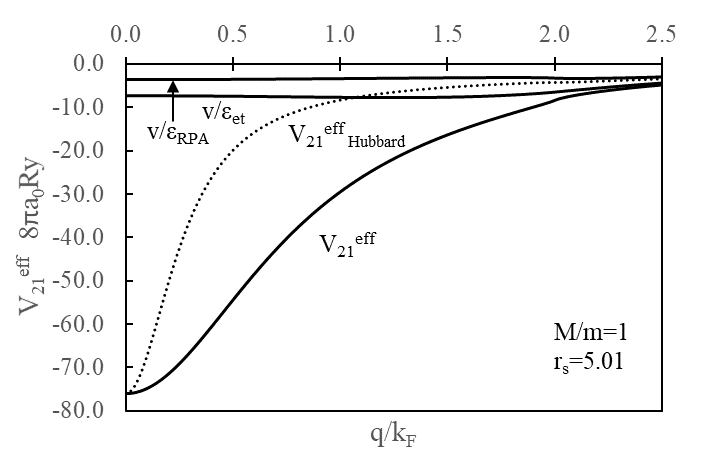}
    \caption{The effective attractive interaction between electrons and positive fermions as a function of wave vector at mass ratio $M/m=1$ for which the equilibrium density is $r_s = 4.176$. The effective interaction from the simple model that correctly includes exchange and correlation is compared to the Hubbard approximation, the coulomb potential divided by the electron-test charge dielectric function in a uniform background, and the RPA for the electron-positive fermion gas. Comparisons are at $80\%$, $100\%$ and $120\%$ of the equilibrium $r_s$.}
    \label{fig5}
\end{figure}
\label{fig5}
Figure \ref{fig5} shows that the electron-test charge (similar to Thomas-Fermi) and RPA interactions are poor approximations and should not be used. The Hubbard approximation agrees with the current result only at $q=0$ and seriously underestimates the strength of the interaction at intermediate $q$. The correct expression for $V_{21}^{\text{eff}}$ is as simple as the Hubbard approximation and is readily available for use in other calculations. Furthermore, the Hubbard interaction completely misses the charge density waves that occur at mass ratios greater than $4.97$.

For $M/m=1$, the strength of the interaction $V_{21}^{\text{eff}}$ increases with $r_s$ and diverges at the compressibility instability. The equilibrium $r_s$ is $80 \% $ of the $r_s$ at the compressibility instability at $q=0$, as was shown in Paper I \cite{ref1}.

\begin{figure}
    \centering
    \includegraphics[width=1\linewidth]{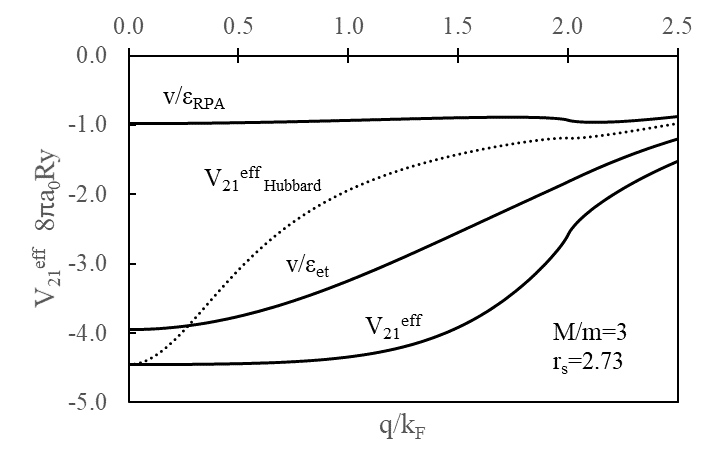}
    \includegraphics[width=1\linewidth]{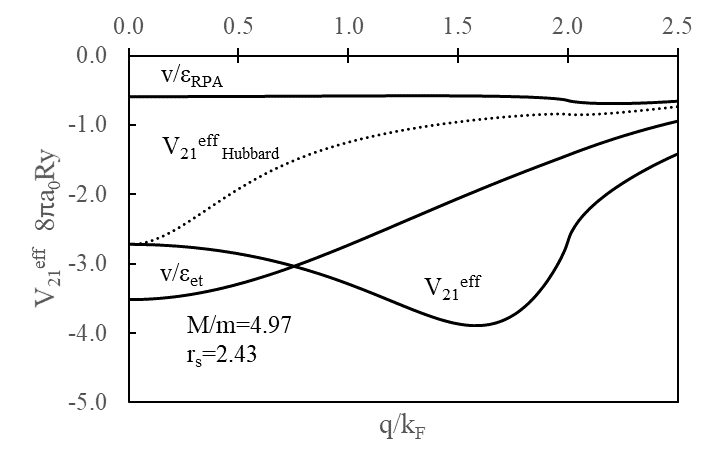}
    \includegraphics[width=1\linewidth]{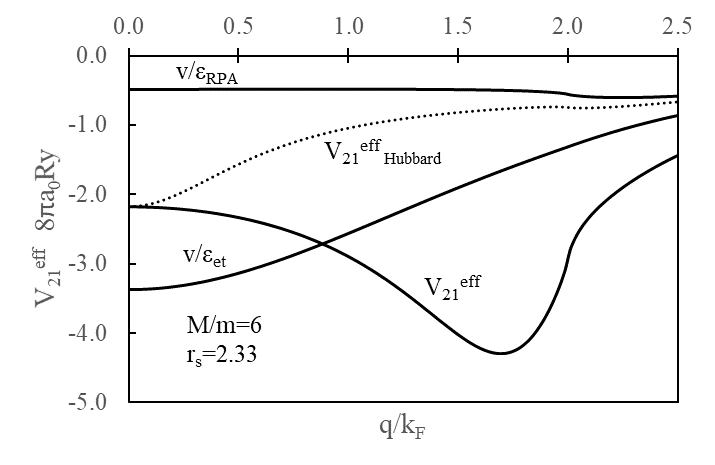}
    \includegraphics[width=1\linewidth]{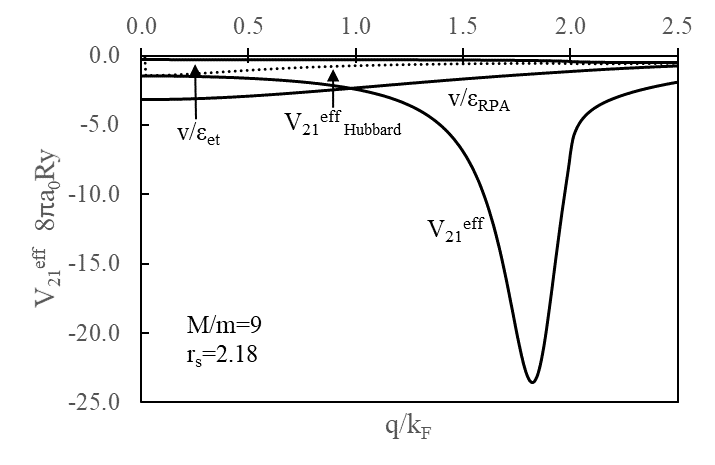}
    \caption{The effective attractive electron-positive fermion interaction versus wave vector. The effective interaction is plotted at the equilibrium $r_s$ where the energy is a minimum for each mass ratio. The different interactions are described in the caption for Fig. \ref{fig5}. }
    \label{fig6}
\end{figure}

I have calculated the same interactions in Fig. \ref{fig5} for different mass ratios each at their equilibrium $r_s$ and these are shown in Fig. \ref{fig6}.

At all mass ratios, there is an instability at $q=0$ at different $r_s$ for each mass ratio. This indicates a condensation or phase separation and is referred to as the compressibility instability because the bulk modulus goes to zero. For mass ratios greater or equal to $4.97$, there is an additional instability at finite $q$ (charge density wave) at a different $r_s$ which smaller than the $r_s$ of the $q=0$ instability. However $r_s$ of the charge density wave is higher than the equilibrium $r_s$ for mass ratios less than 9.

The simple model predicts that the electron-positive fermion gas at the equilibrium $r_s$ is stable for mass ratios less than 9. Although the model breaks down when the charge density wave occurs, it may continue to be accurate if the amplitude of the charge density wave is small. See the discussion in Appendix A. 

Figure \ref{fig6} shows that at small mass ratios $M/m$, the interaction $V_{21}^{\text{eff}}$ is most attractive at small wave vector, and falls off with increasing wave vector. At $M/m=3$, $V_{21}^{\text{eff}}$ remains nearly constant until approximately $q/k_F=1.5$. At $M/m=4.97$, $V_{21}^{\text{eff}}$ shows a pronounced dip near $q/k_F=1.8$ which reflects the incipient charge density wave. This effect is even more pronounced at $M/m=6$. At $M/m=9$, the equilibrium $r_s$ is quite close to the critical $r_s$ for the charge density wave and the enhancement of the interaction is dramatic.

In all cases, $V_{21}^{\text{eff}}$ is substantially more attractive than the Hubbard approximation at intermediate and large wave vector. This means that there will be more large $q$ scattering of electrons from the positive fermions (and vice versa) which will affect the superconducting and transport properties which are discussed in Sections VII and VIII.

\section{ELECTRON-ELECTRON INTERACTION}
The objective is to calculate the electron-electron interaction needed for superconductivity and for normal state electron-electron scattering that contributes to transport properties. Consider two {\it particular} electrons that are interacting via the screened effective potential generated by the other electrons and positive fermions in the gas. The exchange and correlation interactions amongst electrons and positive fermions in the gas are treated in a mean field approximation and these particles are called {\it average} to distinguish them from the two particular electrons that are scattering. The two {\it particular} electrons must have a two-electron state that is antisymmetric with
respect to their interchange, and correlation is taken into account by calculating multiple scattering beyond the Born approximation.

One {\it particular} electron in the Fermi sea is considered the perturbation that induces the screened response of the {\it average} electrons-and {\it average} positive fermions in the gas wich results in the effective potential that the other {\it particular} electron feels. The interactions between parallel and antiparallel electrons are different.

This is the basic physics of the intuitive Kukkonen-Overhauser \cite{ref3} mean field approach which was shown by Kukkonen and Wilkins  \cite{ref18} and Vignale and Singwi \cite{ref20}  to be equivalent to summing a subset of Feynman diagrams.

I follow the derivation for the electron-electron interaction in the uniform electron gas described in detail in Appendix B of Paper I, and partially repeated here.

The first issue is to understand what the disturbance or “external potential” seen by the {\it average} electrons and positive fermions in their Fermi seas in order to calculate the induced densities. When we consider a test charge as the external potential, it is viewed as an infinitesimal charge at a fixed position that interacts with the Fermi sea only through the coulomb potential. The fixed position implies that the test charge has infinite mass and does not react in any way to the electrons in the Fermi sea. The infinitesimal charge justifies linear response theory. An electron has the same coulomb interaction as a test charge but has unit charge. However, this electron is also identical to other electrons with the same spin and has finite mass. This leads to exchange and correlation interactions with electrons of the same spin, and a correlation interaction with electrons of opposite spin, and a correlation interaction with the positive fermions. Although the charge of the electron is not infinitesimal, linear response theory is still used.

The relative spin of the perturbing electron and an {\it average} electron in the Fermi sea must now both be specified. The effective fields felt by {\it average} electrons and the {\it average} positive fermions are used to calculate the induced densities in the electrons and positive fermions.

Lastly, the wave function of the two {\it particular} interacting electrons must be antisymmetric, and cannot be treated in the same way as an {\it average} electron.

The general equations defining the effective interactions are derived in Appendices B \& C of Paper I. It is instructive to start with the following two equations that describe the effective interaction of an {\it average} electron to the disturbance of a {\it particular} electron with spin up.
\begin{eqnarray}
V_{2+}^{\text{eff}} &=& \frac{V_{2+}^{\text{ext}} + V_{21}^b (1- 2G_{21}) \Delta n_{1+}}{\varepsilon_{\text{et}}} \; .        \\
V_{2-}^{\text{eff}} &=& \frac{V_{2-}^{\text{ext}}}{ 1 - G_{2-} v \Pi^0_2 }  \label{eq9}
\end{eqnarray}
Note that Eq. \eqref{eq9} does not depend on the positive fermions.      

Although it is often convenient to consider the sum and differences, $V_{+}^{\text{eff}} = (V_{\uparrow}^{\text{eff}} + V_{\downarrow}^{\text{eff}})/2$, $V_+^{\text{ext}} = (V_\uparrow ^{\text{ext}} + V_\downarrow^{\text{ext}})/2$, $\Delta n_+ = \Delta n_\uparrow  + \Delta n_\downarrow$, $G_+ = G_{\uparrow \uparrow} + G_{\uparrow \downarrow}$, the effective interactions we are looking for are the spin dependent versions, $V_{2 \uparrow \uparrow}^{\text{eff}} = V_{2+}^{\text{eff}} + V_{2-}^{\text{eff}}$ and $V_{2 \downarrow \uparrow}^{\text{eff}} = V_{2+}^{\text{eff}} - V_{2-}^{\text{eff}}$.

If the disturbance is an electron with spin-up, the external potential seen by an {\it average} electron with spin-up is $v \rho_{2\uparrow}(1- 2G_{2\uparrow \uparrow})$ and $v \rho_{2\uparrow}(1- 2G_{2\downarrow \uparrow})$ is the external potential felt by an {\it average} electron with spin-down. The disturbance seen by the positive fermions is $v \rho_{2\uparrow}(1- 2G_{12})$. The resulting external potentials are
\begin{eqnarray}
V_{2\uparrow \uparrow}^{\text{ext}} &=& \frac{1}{2} \left( v \rho_{2\uparrow}(1- 2G_{2\uparrow \uparrow}) + v \rho_{2\uparrow}(1 - 2G_{2\downarrow \uparrow}) \right) \nonumber  \\
 &=& v \rho_{2\uparrow}(1- G_{2+})      ,  \\
V_{2 \downarrow \uparrow}^{\text{ext}} &=&  \frac{1}{2} \left( v \rho_{2\uparrow}(1- 2G_{2\uparrow \uparrow}) – v \rho_{2\uparrow}(1- 2G_{2\downarrow \uparrow}) \right) \nonumber \\
&=& - v \rho_{2\uparrow}G_{2-}   \; ,        \\
V_{1+}^{\text{ext}} &=& v \rho_{2\uparrow}(1- 2G_{12}) \; \; .
\end{eqnarray}
In order to calculate electron-electron matrix elements that are needed for superconductivity and the electron-electron scattering contribution to the electrical and thermal resistivities, another step is needed. The total wave function for two electrons must be antisymmetric. This incorporates exchange in the correct way. The average effective potential between two parallel electrons in the Fermi sea already includes exchange and correlation in an average (local mean field) sense. Similarly the effective potential between two opposite spin electrons includes correlation. Using these potentials for matrix elements will double count exchange and correlation between these two particular electrons.

The effective potentials needed for matrix elements and scattering amplitudes are obtained by removing the average direct exchange and correlation between parallel spin electrons, and removing the average direct correlation between opposite spin electrons. The term $-v \rho_{2\uparrow} 2G_{2\uparrow \uparrow}$ is subtracted from $V_{2\uparrow \uparrow}^{\text{eff}}$ and $–v \rho_{2\uparrow}2G_{2 \downarrow \uparrow}$ from $V_{2\downarrow \uparrow}^{\text{eff}}$.

The effective potential for matrix elements or scattering of parallel and antiparallel electrons are denoted by $V^{\text{scat}}_{2\uparrow \uparrow}$ and $V^{\text{scat}}_{2\downarrow \uparrow}$.
\begin{eqnarray}
V^{\text{scat}}_{2\uparrow \uparrow} &=& v \rho_{2\uparrow} \! \left[ \frac{1 + (1 - G_{2+}) G_{2+} v \Pi^0_2}{ 1 + v (1 - G_{2+}) \Pi^0_2 }  \right. \label{eq13} \\
&& \hspace{1cm} \left.  + \frac{ G_{2-}^2 v \Pi^0_2}{1 – v G_{2-} \Pi^0_2} \right] + \frac{V_{21}^b (1- 2G_{21}) \Delta n_{1+}}{\varepsilon_{\text{et}}}     \nonumber \\
V^{\text{scat}}_{2\downarrow \uparrow} &=& v \rho_{2\uparrow} \! \left[ \frac{1 + (1- G_{2+}) G_+ v \Pi^0_2}{1 + v(1-G_{2+}) \Pi^0_2}    \right. \label{eq14}  \\
&& \hspace{1cm} \left. -  \frac{G_{2-}^2 v \Pi^0_2}{1 – v G_{2-} \Pi^0_2} \right] + \frac{V_{21}^b (1- 2G_{21}) \Delta n_{1+}}{\varepsilon_{\text{et}}}  \nonumber
\end{eqnarray}

These scattering potentials  are exactly the same as the scattering potentials for the previously known electron-electron electron scattering in a uniform background Eqs. (B15-16) of Ref. \cite{ref1} plus an additional term proportional to the induced density of positive fermions, which is given by  $\Delta n_{1+} = -\Pi^0_1 V_{1+}^{\text{eff}}$ and $V_{1+}^{\text{eff}} = v \rho_{2 \uparrow}(1- 2G_{12}) / \Delta $  in analogy to Eq. \eqref{eq7}. Assuming the bare interaction is the coulomb interaction, and $G_{12} = G_{21}$, the attractive term that leads to superconductivity can be written as
\begin{eqnarray}\label{eq15}
U_{21} \! = \! \frac{ V_{21}^b (1- 2G_{21}) \Delta n_{1+}}{ \varepsilon_{\text{et}}} \!=\! \frac{ -v \rho_{2\uparrow}(1-2G_{21})^2 v\Pi^0_1}{ \varepsilon_{\text{et}} \Delta} \, . 
\end{eqnarray}

A brief discussion of these effective potentials is in order. The Kukkonen Overhauser mean local field approach is simple and intuitive, but is an approximation. Kukkonen and Wilkins \cite{ref18}, Vignale and Singwi \cite{ref20}, and the excellent textbook by Giuliani and Vignale \cite{ref21}  use formal methods and Feynman diagrams to calculate the four-point scattering function.  They also used the local approximation that allowed a subset of diagrams to be summed. The remaining diagrams were ignored. Fortunately these two approaches agree, but they remain approximations without any certainty as to their accuracy. These interactions are expressed in terms of local field factors that must satisfy sum rules at small and large wave vectors and zero frequency. 

Also fortunately, Quantum Monte Carlo calculations have advanced significantly and have been used to calculate the response functions of the electron gas and extract the wave vector dependence of the local field factors that is required by the theoretical dielectric linear response formulation of the electron gas problem. These calculations verify the compressibility sum rule and show that the density and spin local field factors rise as $q^2$ from $q=0$ up to nearly $q= 2k_F$. The underlying theory of the local field factor at intermediate wave vector is not known and the behavior is empirically taken from connection with the Quantum Monte Carlo calculations. See Refs. \cite{ref13,ref19} and further references therein. However, this important $q^2$ behavior leads to the charge density waves discussed in this paper. The $q^2$ behavior is equivalent to the local spin density approximation in density function theory \cite{ref21}. The limitations of this assumption are discussed in Appendix A. The frequency dependence of the local field factor is also relatively unknown, and ignored in this paper.

Superconductivity and electron-electron scattering calculations in the many body electron gas require a distinction between two particular electrons that are interacting and the rest of the average electrons in the Fermi sea that provide the vertex functions and screening of the effective interaction. The two particular electrons must be treated in a two-particle wave function that is antisymmetric under exchange of the two electrons. In the effective interaction, the electron-electron interactions in the Fermi sea are treated in the local approximation using the local field factors.

The antisymmetric two-electron wave functions are a product of a two-particle spatial wave function and a two-particle spin wave function. Two initial state electrons represented by $\psi_1(r_1,p_1)$ with spin $s_1$ and  $\psi_2(r_2, p_2)$ with spin $s_2$ interact through the effective interaction to become final states with momenta $p_1 – q$ and $p_2 + q$. The two spins can have four orientations $\braket{\uparrow \uparrow}$, $\braket{\downarrow \downarrow}$, $\braket{\uparrow \downarrow}$ and $\braket{\downarrow \uparrow}$. If the spin part of the wave function is antisymmetric, then the spatial part must be symmetric and vice versa. This leads to four possibilities for the two electron states.

Three possibilities called triplet states have an antisymmetric spatial part $1/2^{1/2} (\psi_1(r_1, p_1) \psi_2(r_2, p_2) – \psi_2(r_1, p_1) \psi_1(r_2, p_2))$ multiplied by one of the symmetric spin wave functions,  $\braket{\uparrow \uparrow}$, $1/2^{1/2}( \braket{\uparrow \downarrow} + \braket{\downarrow \uparrow })$ or $\braket{\downarrow \downarrow }$. These have total spin $=1$ with $z$ component $+1$, $0$ and $-1$.

The remaining two state wave function is the singlet state that has a symmetric spatial part $1/2^{1/2} ( \psi_1(r_1, p_1) \psi_2(r_2, p_2) + \psi_2(r_1, p_1) \psi_1(r_2, p_2)$ and antisymmetric spin wave function $1/2^{1/2}( \braket{\uparrow \downarrow} - \braket{\downarrow \uparrow})$ or $\braket{\downarrow \downarrow}$. The triplet state has total spin $S=0$ and $z$ component $0$.

Electron-electron scattering includes all of the two electron states. Ordinary BCS superconductivity is in the singlet state, but p-wave superconductivity is in the triplet state. The simplest approximation is that the electrons are plane waves that are used to calculate matrix elements with the effective interaction. This is equivalent to the first Born approximation which is known to overestimate the effective scattering \cite{ref17}. See the discussion in Section VIII. The triplet state with $z$ component spin equal to zero, and the singlet state are examples of entangled states.

In order to use the mechanics of spin manipulation based on the Pauli matrices, it is convenient to write the effective interaction in a spin invariant form.
\begin{eqnarray}\label{eq16}
V_{\text{ee}}^{\text{eff}} ( \sigma_1 \, , \sigma_2 \, , \, \sigma_q \, , \,  \omega ) = V_0 + J \sigma_1 \cdot  \sigma_2 + U_{21} \;,
\end{eqnarray}
where the spin matrix element of $\sigma_1 \cdot \sigma_2 = -3$ in the singlet state and $+1$ in the triplet state.

From equations (\ref{eq13}-\ref{eq15}), $V_0$, $J$ and $U_{21}$ are given by
\begin{eqnarray}
V_0 &=& \frac{4 \pi e^2}{q^2} \left[\frac{1 + (1- G_{2+}) G_{2+} v \Pi^0_2}{ 1 + v(1-G_{2+}) \Pi^0_2} \right] \; ,\label{eq17} \\
J &=&  -\frac{4 \pi e^2}{q^2} \frac{ G_{2-}^2 v \Pi^0_2}{ 1 – v G_{2-} \Pi^0_2} \; , \label{eq18}\\
U_{21} &=& - \frac{4 \pi e^2}{q^2} (1-2G_{21})^2 \frac{ v\Pi^0_1}{\varepsilon_{\text{et}} \Delta} \; . \label{eq19}
\end{eqnarray}

The first two terms in Eq. \eqref{eq16} are exactly the same effective electron-electron interaction calculated in a uniform positive background. Note that the spin independent term $V_0$ is positive (repulsive) while the second term which is spin dependent is also positive (repulsive) with a coefficient of $3$ for the singlet state, but the second term is negative (attractive) with a coefficient of $1$ for each of the triplet states. The term $U_{21}$ results from the additional screening due to the positive fermions and is always attractive.

Equations (\ref{eq16}-\ref{eq19}) are the same as Eqs. (12-13) of Vignale and Singwi \cite{ref6}. Their notation is different and some algebra is required to show the equivalence. Giuliani and Vignale \cite{ref21} show that in the limit of the positive fermions having much higher mass than the electrons, $U_{21}$ in Eq. \eqref{eq19} reduces to the well-known electron-phonon interaction as shown earlier by Kukkonen and Overhauser \cite{ref3}.

None of the physics reproduced above is new, but is included so that the reader can understand the effects of using current local field factors that predict charge density waves in these general equations, compared to the Hubbard approximation that was used previously. The new and old results are compared in the figures below.

The positive fermion-positive fermion interactions are the same as Eqs. (\ref{eq13}-\ref{eq15}) with the indices $1$ and $2$ interchanged.

\begin{figure*}[!ht]
    \begin{center}
    \includegraphics[width=0.39\linewidth]{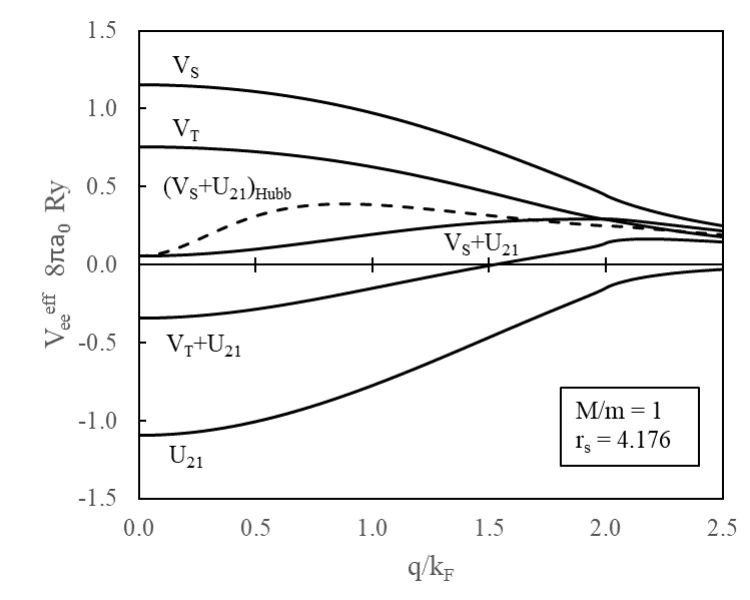}
    \includegraphics[width=0.39\linewidth]{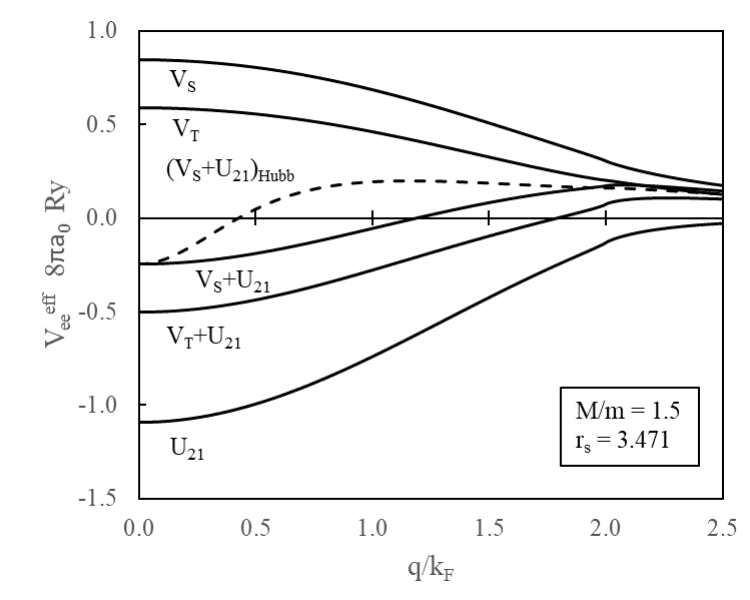}
    \includegraphics[width=0.39\linewidth]{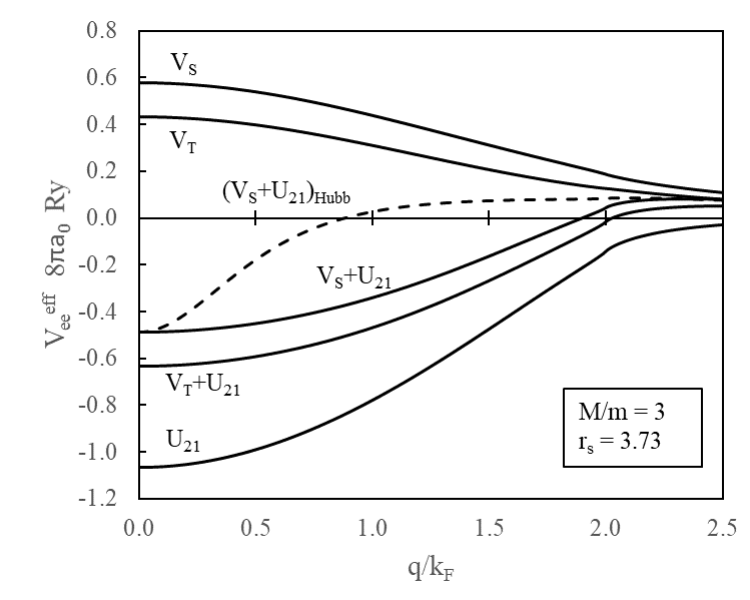}
    \includegraphics[width=0.39\linewidth]{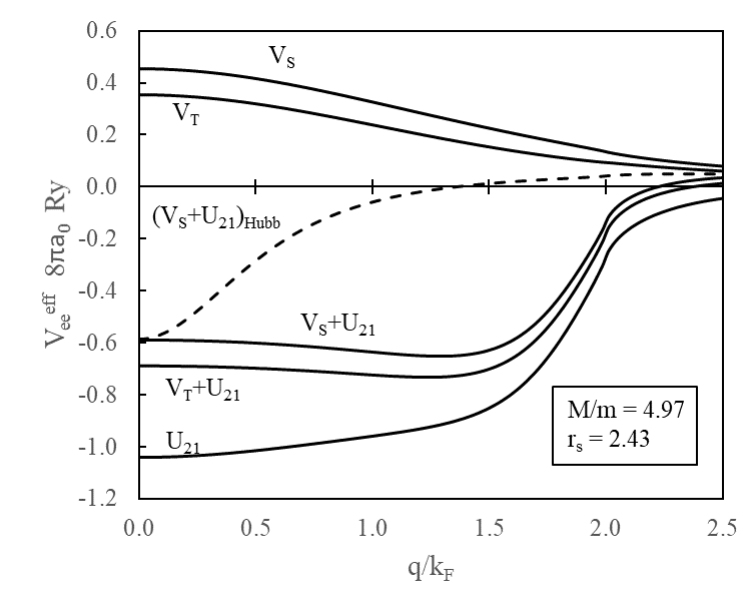}
    \includegraphics[width=0.39\linewidth]{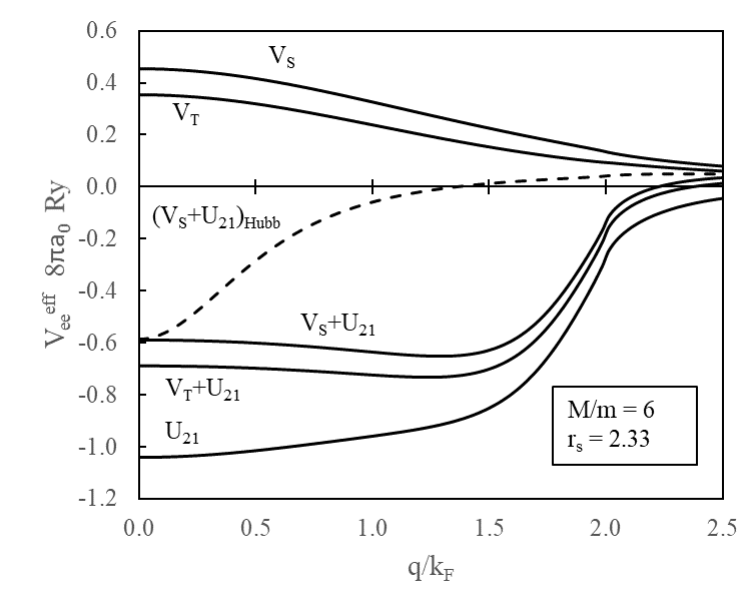}
    \includegraphics[width=0.39\linewidth]{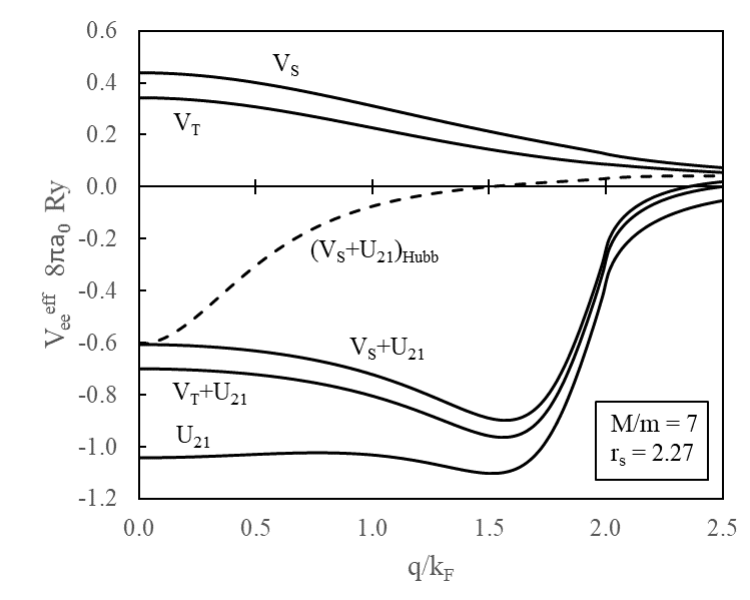}
    \includegraphics[width=0.39\linewidth]{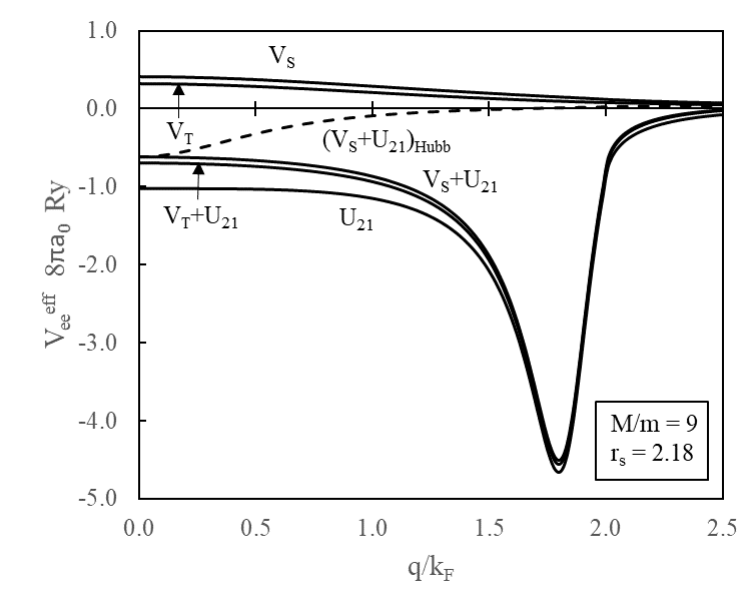}
    \end{center}
    \caption{The effective electron-electron interaction $V_{\text{ee}}^{\text{eff}}$ using the current local field factors is
compared to the Hubbard approximation for different mass ratios $M/m$ of the positive fermions to the electrons. The density $r_s$ is chosen to be the unique equilibrium density for each mass ratio. The repulsive portion of the interaction is shown for both the singlet $V_S$ and triplet $V_T$ states.
The positive fermion mediated attractive term $U_{21}$ is also shown, along with the total interaction $V_{\text{ee}}^{\text{eff}}$ for both the singlet and triplet states. The Hubbard interaction is shown for the singlet state
only.}
    \label{fig7}
\end{figure*}

The different contributions to the effective electron-electron potential $V_{\text{ee}}^{\text{eff}}$ given in Eqs. (\ref{eq16}-\ref{eq19}) are shown in Fig. \ref{fig7}. The singlet $(V_S=V_0-3J)$ and triplet $(V_T=V_0+J)$ interactions are shown and the interaction using current local field factors are compared with the Hubbard approximation.

The first observation from Fig. \ref{fig7} is that the Hubbard local field factor yields a poor approximation even though it agrees with the current results  at $q=0$ and at $q > 2k_F$.  The current results using a local field factor that increases as $q^2$ are dramatically different particularly as the mass ratio increases. This means that the current local field factors will predict much larger superconducting transition temperatures and higher resistivity in the normal state.

The second observation is simply the obvious fact that the effective repulsive interaction for the triplet state is always smaller than that of the singlet state. The difference is relatively small. Superconductivity calculations have different weighted averages of $V_{\text{ee}}^{\text{eff}}$ in the singlet state which favors small wave vector compared to the triplet state. 

For equal masses $M/m=1$, the effective singlet interaction is purely repulsive which would not predict superconductivity. The triplet state is slightly attractive. 

For mass ratios below $4.97$, the only instability is the compressibility instability at $q=0$ and $V_{\text{ee}}^{\text{eff}}$ becomes attractive with a minimum at $q=0$ which should predict singlet superconductivity. At the density of the energy minimum $r_s$ is $80\%$ of the $r_s$ of the compressibility instability. The depth of the attractive effective interaction increases with $r_s$ and gets very large near the instability, but there is no obvious way beyond a pseudo-potential to increase $r_s$. Pressure could be used to decrease $r_s$.
 
At the mass ratio $M/m= 4.97$, a critical point is reached at $r_s = 3.036$ where the compressibility instability at $q=0$, and a charge density wave at $q/k_F = 0.83$ coexist \cite{ref1}. For this mass ratio, the equilibrium density is at $r_s = 2.43$ which is not particularly close to either instability, but now the effect of the charge density wave becomes apparent with $V_{\text{ee}}^{\text{eff}}$ being flatter and more attractive at larger wave vectors. This increases the difference with the Hubbard approximation and will lead to higher predictions of the superconducting transition temperature for both singlet and triplet states.

As shown in Ref. \cite{ref1}, as the mass ratio increases, the wave vector of the charge density wave increases and the $r_s$ where the charge density wave first appears becomes smaller than the $r_s$ of the compressibility instability at $q=0$. See Fig. 7 of \cite{ref1} where the $r_s$ of the energy minimum, compressibility instability and the charge density wave instability are compared. This occurs at $M/m > 9$. I do not make calculations at larger mass ratios. See discussion in Appendix A.

The effective interactions $V_{\text{ee}}^{\text{eff}}$ for mass ratios $M/m = 6\,  7  \, \& \,  9$ are also shown in Fig. \ref{fig7}. At these mass ratios and their equilibrium densities, there are no compressibility or charge density wave instabilities. However, as stated above, the $r_s$ of the compressibility instability is becoming closer to that of the equilibrium density and therefore the influence of the charge density wave is becoming much more significant. The attractive minimum of $V_{\text{ee}}^{\text{eff}}$ is no longer at $q=0$, but appears near the wave vector of the incipient charge density wave and is much more attractive. This implies even higher singlet superconducting transition temperatures. In calculating scattering, there is much more large wave vector scattering which will lead to much higher normal state electrical resistivity.

\section{POSITIVE FERMION-POSITIVE FERMION INTERACTION}
The calculation of the effective interaction between positive fermions is exactly the same as for electrons. The physical equations only depend on charge squared so that they may be used directly. The only difference is the mass of the positive fermions compared to the electrons $M/m$. Equations (\ref{eq17}-\ref{eq19}) may be used with the interchange of subscripts $1$ (positive fermions) and $2$ (electrons) and the letters $\text{e}$ for electrons and h which I use for the positive fermions or holes.

\begin{figure}
    \centering
    \includegraphics[width=1\linewidth]{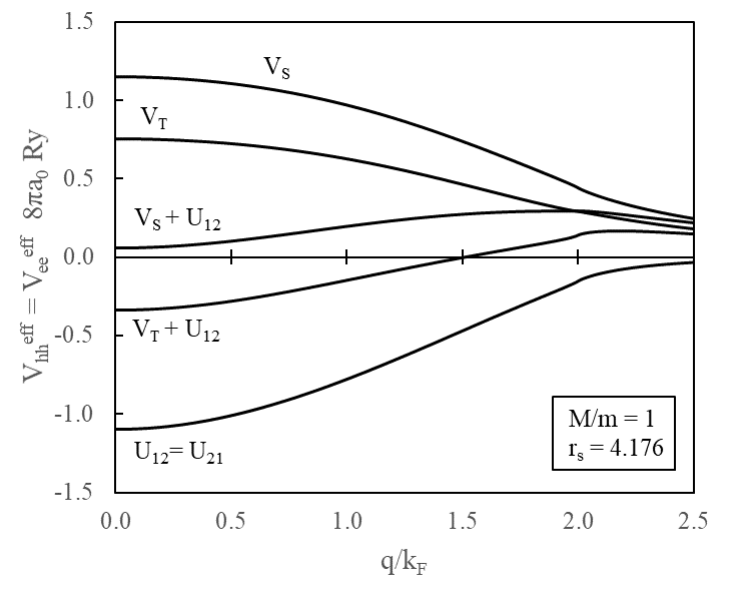}
    \includegraphics[width=1\linewidth]{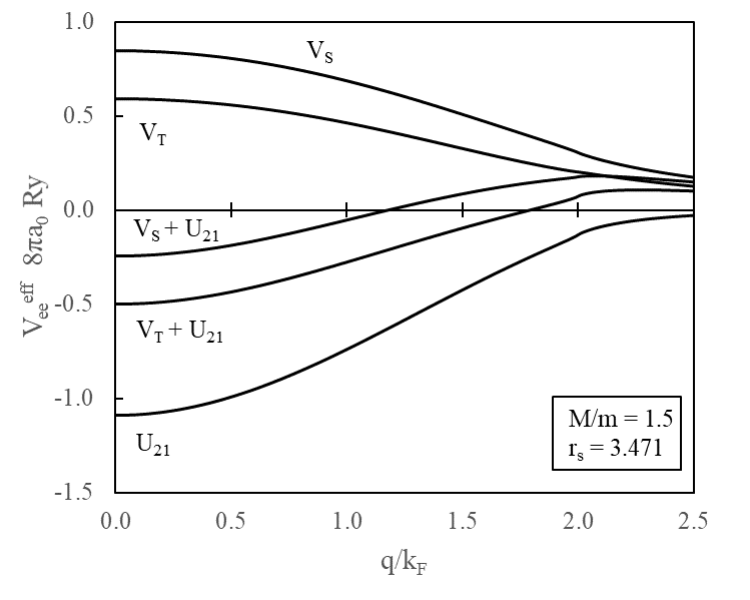}
    \includegraphics[width=1\linewidth]{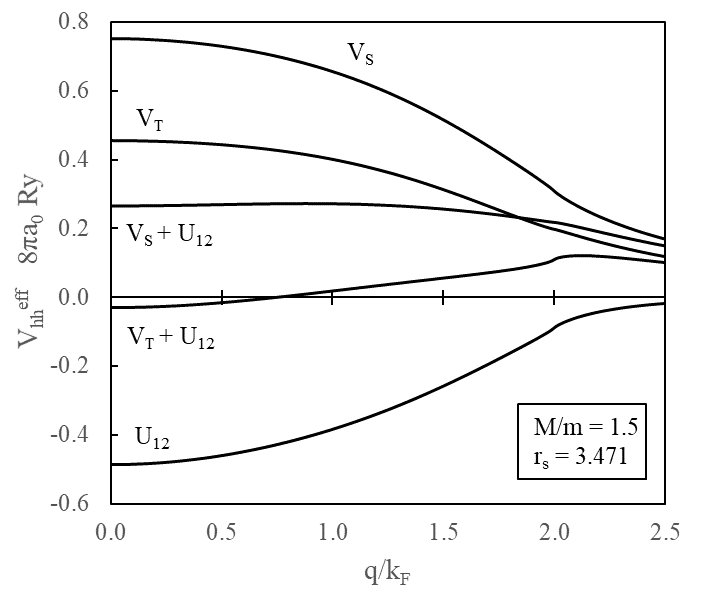}
    \caption{Comparison of the positive fermion-positive fermion effective interaction to the electron-electron interaction at mass ratio $M/m =1$ where both interactions are equal and at $M/m = 1.5$, where they are different.}
    \label{fig8}
\end{figure}

The repulsive portion of the positive fermion-positive fermion interaction is modestly dependent on mass and becomes less repulsive than the electron-electron interaction as $M/m$ increases. However the attractive term is much more sensitive and gets weaker for the holes by the ratio $(m/M)^2$ at $q=0$. The two interactions are directly compared in Fig. \ref{fig8}.

Using Eq. \eqref{eq7}, the ratio of the attractive terms for the electrons, compared to the positive fermions is given by
\begin{eqnarray}
\frac{U_{21}}{U_{12}} = \frac{\frac{\Delta n_{1+}}{\varepsilon_{\text{et}}}}{ \frac{\Delta n_{2+}}{ \varepsilon_{\text{ht}}}} \;=\;  \frac{ \Pi^0_1}{\Pi^0_2} \cdot  \frac{ \varepsilon_{\text{ht}}}{ \varepsilon_{\text{et}}}
\end{eqnarray}
At $q=0$, this ratio becomes $(\Pi^0_1/ \Pi^0_2)^2 = (M/m)^2$. The Lindhard polarizability at $q=0$ is the density of states at the Fermi surface $N(0) = m k_F/\pi^2 \hbar^2 = q_{TF}^2/4\pi e^2$, which shows the scaling with mass and the relationship between the density of states and the Thomas Fermi screening wave vector.
Figure \ref{fig8} shows that the only potential attractive interaction between two positive fermions is for triplet pairing and mass ratios $M/m< 1.5$. If such a system existed, it is might be possible to see pairing or superconductivity in both the electrons and positive fermions.

For $M/m > 1.5$, there is no net attraction for the positive fermions. Note that I have defined the positive fermions as the heavier particle. If the positive fermions had a lower mass than the electrons, the results would be the same, but the roles of the electrons and positive fermions are reversed.

\section{Magnetic response.}
Applying an external magnetic field H (in the z direction) to a paramagnet induces a magnetization M in the same direction of the external field that adds to produce a local field that is larger than the applied field.
\begin{eqnarray}\label{eq21}
B = \mu_0(H +M)
\end{eqnarray}
where $\mu_0$ is the vacuum permeability.

For a two component system $M = M_1 + M_2$.  Linear response theory connects the magnetization to the local field. In the simple model of this paper, the electrons and positive fermions do not have a magnetic interaction. In this case, the electrons and positive fermions respond independently, $M_1 = \chi_1 B$ and $M_2 = \chi_2 B$. This yields
\begin{eqnarray}\label{eq22}
B = \frac{ \mu_0 H }{  1 – (\chi_1 + \chi_2)} \; .            
\end{eqnarray}

If $(\chi_1 + \chi_2) \geq 1$, the system will be unstable. If the instability occurs at $q=0$, it could be a ferromagnetic transition. At finite $q$, it signals a spin density wave.

The $q$ dependent Pauli susceptibility is given by $\chi_0 = \mu_B^2 \Pi^0_2(q)$, where the Bohr magneton is given by $\mu_B = e \hbar / 2m $ and the Lindhard free electron polarizability is $\Pi^0_2(q) = (m k_F/\pi^2 \hbar^2) L(q)$. $L(q)$ is the dimensionless Lindhard function \cite{ref14,ref21} that equals $1$ at $q=0$, remains near one until it begins to fall off quickly near $q =2 k_F$.
For the electron mass, $\chi_{02} = 2.59/r_s \cdot 10^{-6}$ at $q=0$. For the positive fermion, $\chi_{01} = m/M \cdot  \chi_{02}$. In this simple model, the Pauli susceptibility enhancement factors for the electrons and the positive fermions have the same form: 

\begin{eqnarray}
\chi_2 = \frac{\chi_{02}}{ 1 - G_{2-}v\Pi^0_2 } \; , \label{eq22}      \\
\chi_1 = \frac{ \chi_{01}}{ 1 - G_{1-}v \Pi^0_1} \; . \label{eq23}
\end{eqnarray}
Quantum Monte Carlo calculations of $\chi_2$ have confirmed that the enhancement factor and therefore the local field factor is precisely consistent with the susceptibility sum rule derived by differentiating the energy. Quantum Monte Carlo calculations have also shown that the initial $q^2$ wave vector dependence of the spin local field factor $G_-$ continues until nearly $q=2 k_F$, as it does for the density local field factor $G_+$, however there is nothing equivalent to the additional screening by the positive fermions, so no spin density waves occur. Other Quantum Monte Carlo calculations at large $r_s =120$ indicate that there is no magnetic phase transition at q=0. See Ref. \cite{ref13} and further references therein.

\begin{figure}
    \centering
    \includegraphics[width=1\linewidth]{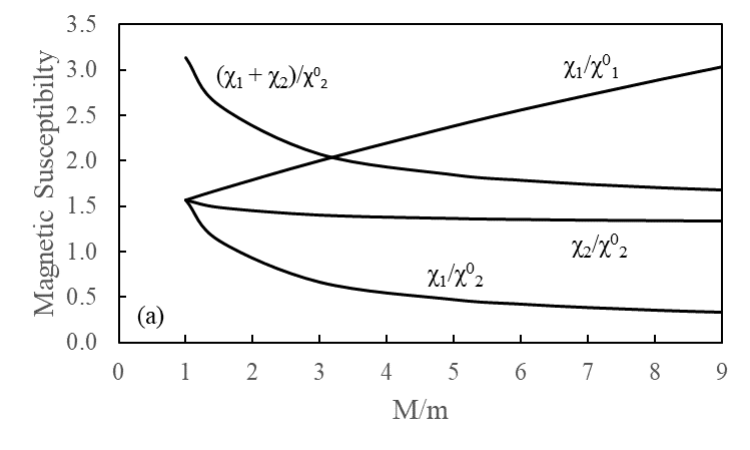}
    \includegraphics[width=1\linewidth]{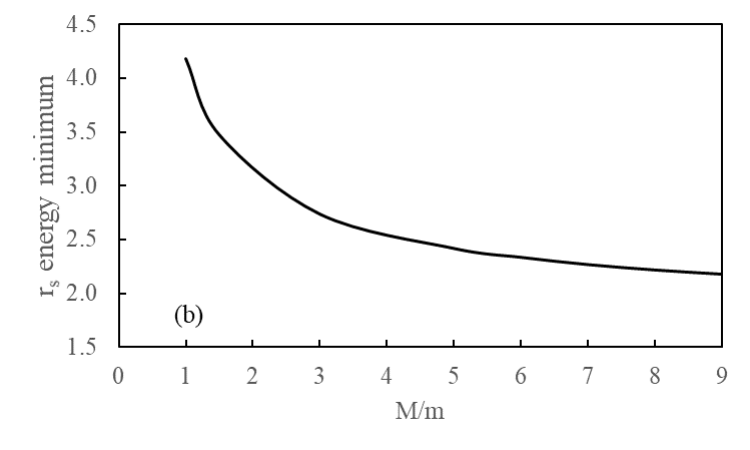}
    \caption{Magnetic susceptibility enhancement of the electron-positive fermion gas in the simple model. The susceptibility enhancement is the ratio of the calculated susceptibility to the susceptibility of the free electron (or positive fermion) gas. The two gases respond independently
because they are not magnetically coupled. The susceptibility enhancement is shown in (a) for different mass ratios. For each mass ratio, there is a unique $r_s$ where the energy is a minimum and this is where the susceptibility is calculated. The energy minimum $r_s$ for each mass ratio $M/m$ is
shown in (b).}
    \label{fig9}
\end{figure}

The electron-positive fermion gas has a unique density minimum for every mass ratio $M/m$. The total susceptibility is calculated in the simple model for different mass ratios and is displayed in Fig. \ref{fig9}.

Figure \ref{fig9} shows the susceptibility enhancement for mass ratios up to $M/m=9$. Above $M/m=9$, the simple model predicts that the system will have a charge density wave which invalidates the basic assumption of a uniform background in the simple model. The simple model may continue to be useful at larger mass ratios, but I have not investigated this possibility.
It is important to remember that the effective density of the positive fermions is $M/m\cdot r_s$.  The electron gas in a uniform background has been theoretically investigated using Quantum Monte Carlo techniques for $r_s \leq 120$, and no charge or spin density waves have been found. In the electron-positive fermion gas, the electron density at the energy minimum is always at $r_s \leq 4.18$. However the effective $r_s$ of the positive fermions becomes very large for large mass ratios. For this reason, the susceptibility enhancement at large mass ratios may not be accurate. This is another limitation of the simple model.
The predicted susceptibilities are tiny compared to unity and local field enhancements in Eq. (\ref{eq23}-\ref{eq24}) are negligible.

However an extension of the simple model that allows another type of electrons that can exchange with the first type would result in a susceptibility enhancement factor of the form $\chi_2 = \chi_{02}/ (1 - G_{2-}v \Pi^0_2 - G_{3-}v \Pi^0_3)$. This system would have a ferromagnetic divergence at q=0 or a spin density wave at finite $q$, if the sum $G_{2-}v\Pi^0_2 + G_{3 -} v \Pi^0_3$ became greater than $1$. This system could exhibit both charge and spin instabilities, which influence both superconductivity and itinerant magnetism.

\section{Superconductivity.}
The possibility of superconductivity in the electron-hole liquid was investigated in Refs. \cite{ref5,ref7,ref8}. I follow the approach of Vignale and Singwi \cite{ref7} and simply calculate the superconducting parameters $\mu $ and $\lambda $  and compare to their results. Their pioneering work of 1985 calculated the ground state energy using a self-consistent scheme, utilized the compressibility sum rule and assumed the Hubbard approximation for the wave vector dependence of the local field factor that yields the effective electron-electron interaction.

The main difference between this work and Ref. \cite{ref7} is the wave vector dependence of the local field factor that was revealed by subsequent Quantum Monte Carlo Calculations and introduces charge density waves into the electron-positive fermion system \cite{ref1}. Proximity of the density $r_s$ at the energy minimum to the charge density wave at different mass ratios significantly enhances the attractive portion of the effective electron-electron interaction which predicts a significantly higher superconducting transition temperature.

\begin{figure}
    \centering
    \includegraphics[width=1\linewidth]{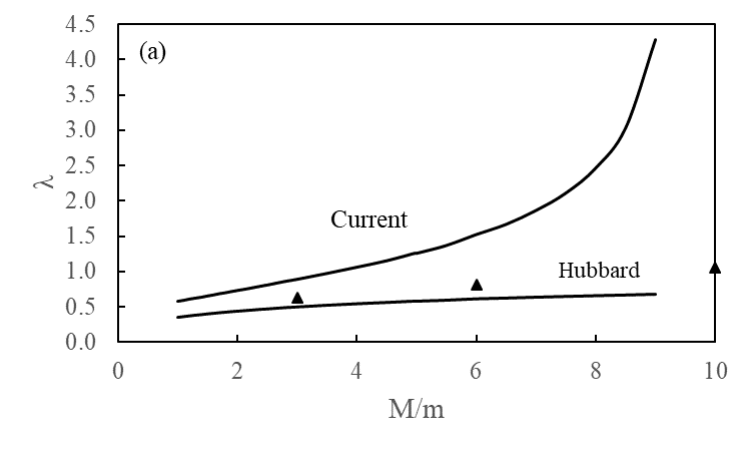}
    \includegraphics[width=1\linewidth]{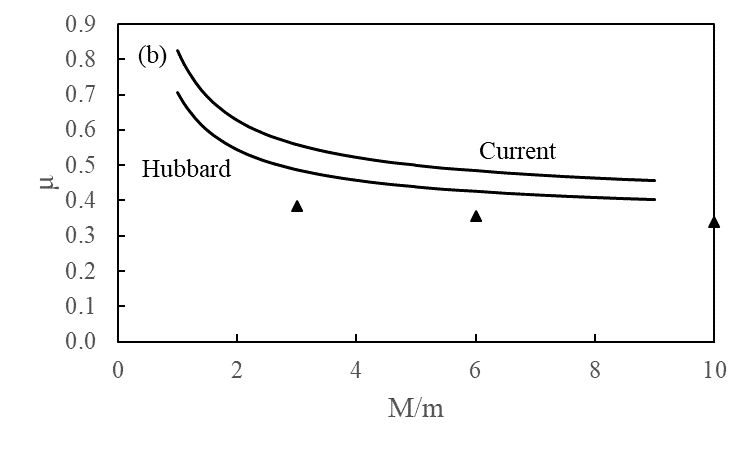}
    \includegraphics[width=1\linewidth]{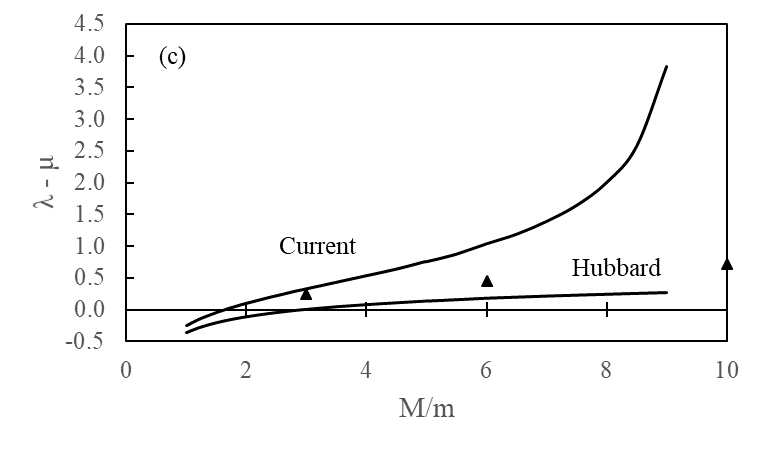}
    \caption{Superconducting parameters $\mu$ and $\lambda$ of the electron- positive fermion gas as a function of mass ratio $M/m$. The calculations are at the density $r_s$ of the unique energy minimum for each mass ratio. The effective electron-electron interaction with current local field factors is compared to the Hubbard approximation used in previous calculations, but using the parameters of the simple model. The triangles are previous results from Vignale and Singwi \cite{ref7} who also used the Hubbard approximation. (a) presents $\lambda$, (b) presents $\mu$, and (c) presents the negative of the net interaction  $\lambda  – \mu$.}
    \label{fig10}
\end{figure}

The effective electron-electron interaction is given in Eqs. (\ref{eq16} - \ref{eq19}). This interaction is wave vector and frequency dependent. The local field factors are taken (without justification) to be independent of frequency. This interaction maybe used in calculating the superconducting transition temperature using more sophisticated methods \cite{ref24}. I am not an expert in superconductivity and simply use the interactions to calculate the superconducting parameters $\mu$ and $\lambda$ to show the effect of using the current interaction compared to the Hubbard approximation. For the singlet state, the effective interaction between two electrons is

\begin{eqnarray}
V_S &=&  V_0 - 3J + U_{21} \; .  
\end{eqnarray}
These equations are given in Eqs. (\ref{eq17}-\ref{eq19}). Note that $J$ is a negative number that increases the repulsion, and that $U_{21}$ is a negative number and this term is attractive. The superconducting parameters are simply averages of this interaction over the Fermi surface.  I follow the approach of Ref. \cite{ref7} that yields
\begin{eqnarray}\label{eq26}
\mu – \lambda &=& \left( \frac{q_{TF}}{2 k_F} \right)^2 \int^{2k_F}_0 dq q (V_0 - 3J + U_{21})\; ,
\end{eqnarray}
where the pre-factor is proportional to the density of states at the Fermi surface. The first two terms produce $\mu$ and the last term $\lambda$. The weighting factor of $q$ in the average over the Fermi surface greater emphasis to large wave vectors and this is where the difference between the Hubbard approximation and the current local field factors is greatest.
For each mass ratio $M/m$ the electron-positive fermion gas has a unique $r_s$ of the energy minimum (see Fig. \ref{fig9}b). Following Ref. \cite{ref7}, I calculate $\mu$ and $\lambda$ at this density and the results are shown in Fig. \ref{fig10}.

The main observation is that attractive superconducting parameter $\lambda$ is large and becomes very large near the charge density wave. The repulsive parameter $\mu$ is also large.

The only difference in Fig. \ref{fig10} between the current calculation and Hubbard approximation results from the value of the local field factor at finite $q$ (see Fig \ref{fig1}). Both agree at q=0 because they both satisfy the compressibility sum rule and initially rise as $q^2$ with the same coefficient. The Hubbard approximation falls below $q^2$ at $q=0.7 k_F$ and is much smaller at $2 k_F$. In contrast, the local field factor derived from Quantum Monte Carlo results continue to rise as $q^2$ to nearly $2k_F$.  This continued $q^2$ behavior is the source of the charge density waves discussed in \cite{ref1} and \cite{ref22}. The calculation of $\mu$ in Fig. \ref{fig10}b is exactly the same as the electron gas in a uniform background. The fact that $\mu$ calculated using the correct local field factor is larger than that of the Hubbard approximation was previously shown analytically in Ref. \cite{ref25} and numerically in Ref. \cite{ref24}. It simply reflects that the effective electron-electron using the current local field factors is more repulsive than that obtained with the Hubbard approximation. The triangles show the results of Vignale and Singwi from 1985 \cite{ref7} who used a similar version of the Hubbard approximation which was the best approximation available at the time. The repulsive superconducting parameter $\mu$ has no unusual behavior and can be calculated at any density because it assumes a uniform positive background.

The situation is more complicated for the attractive superconducting parameter $\lambda$ that is an average of the effective electron-positive fermion interaction, which becomes large and then diverges at the presence of the charge density wave.

The effective interaction treats electrons and positive fermions in exactly the same way. There is not an independent calculation of the effect of phonons as in BCS theory. The additional screening from the positive fermions has contributions from both single particle excitations and collective modes. This is the interaction that should be used in ab initio calculations of superconductivity for the electron-positive fermion gas.

\begin{figure}
    \centering
    \includegraphics[width=1\linewidth]{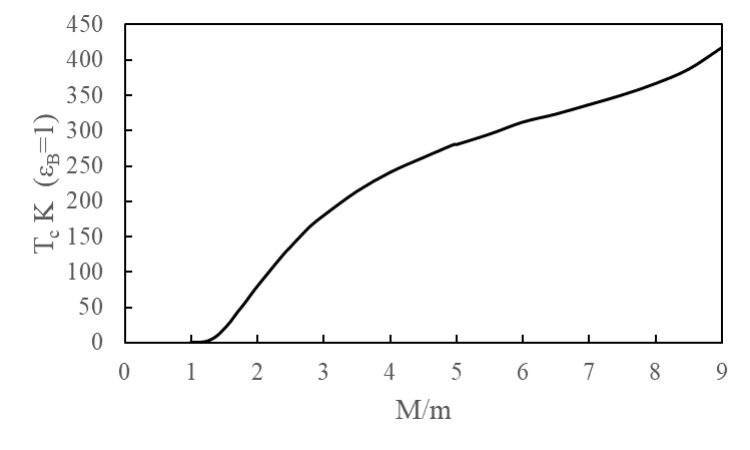}
    \caption{Superconducting transition temperature $T_c$ for s-waves calculated using the modified McMillan formula in Eq. \eqref{eq27} as a function of mass ratio assuming a background dielectric function of unity. The predicted transition temperature would be reduced by $1/ \varepsilon_B^2$. }
    \label{fig11}
\end{figure}

Reference \cite{ref7} used re-normalized $\lambda$ and $\mu$ to calculate the superconducting transition temperature using a McMillan type of formula with the pre-factor that was approximately the Fermi temperature of the positive fermions. I am not sure of the validity of McMillan formula for the strongly coupled electron-positive fermion gas, but it is instructive to examine a version of it shown below.
\begin{eqnarray}\label{eq27}
T_c & = &  T_{F1} \cdot \exp\left( \frac{ -1.04 (1 + \lambda)}{ \lambda – \mu^\ast (1 – 0.62 \lambda )} \right) \; ,     
\end{eqnarray}
Where $T_{F1} = 58200 K/ (M/m \cdot (r_s \varepsilon_B)^2$ is the Fermi temperature of the positive fermions in a background dielectric constant, and $\mu^\ast = \mu/(1+ \mu \log( T_{F2}/T_{F1}) = \mu/(1+ \mu \log(M/m)$ is the renormalization used in Ref. \cite{ref7}. This formula is evaluated at the equilibrium points for each mass ratio and the result is shown in Fig. \ref{fig11}.


Reference \cite{ref7} applied their theory to degenerate semiconductors with background dielectric constants of about $15$ and mass ratios $3-6$, and predicted superconducting transition temperatures of $0.1-1 \, \text{K}$, which is roughly consistent with the predictions of the McMillan formula shown in Fig. \ref{fig11}.

The Fermi temperature of the positive fermions, which is the pre-factor in Eq. \eqref{eq27} falls as their mass increases. However the exponential factor increases faster so that the predicted
superconducting transition temperature shown in Fig. \ref{fig11} increases substantially with mass. This is intuitively obvious when looking at Fig. \ref{fig7} which shows that the effective electron-electron interaction is attractive over a large range of wave vector and becomes increasingly attractive as the mass ratio increases. The predicted transition temperatures of the model system are well above room temperature for mass ratios greater than five. I make no attempt to map the model system onto any real material. 

\begin{figure}
    \centering
    \includegraphics[width=1\linewidth]{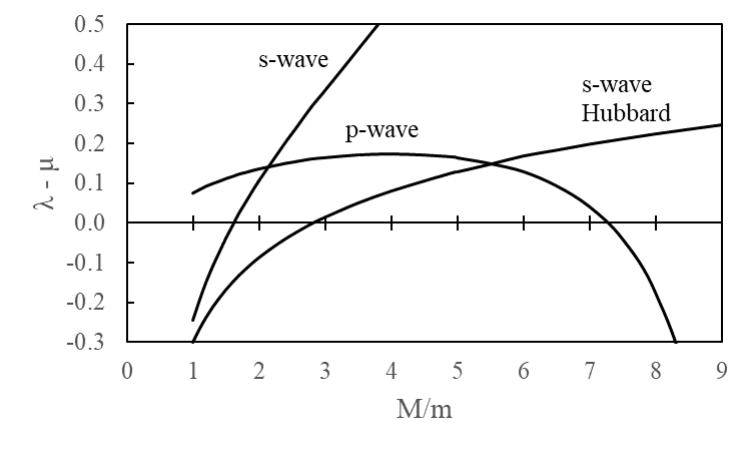}
    \caption{The Fermi surface averaged net attractive electron- electron interaction $\lambda - \mu$ for the electron-positive fermion gas calculated at the density of the energy minimum for different mass ratios $M/m$. The results for antiparallel spins (s-wave) for the correct interaction and the Hubbard approximation are compared to parallel spins (p-wave). }
    \label{fig12}
\end{figure}

The effective electron-electron interaction for $M/m=1$ in Fig. \ref{fig7}, shows that the singlet state is repulsive for all wave vectors, but the triplet state has an attractive interaction at small $q$. For standard BCS theory, the paired electrons are in a singlet state and the Fermi surface averaged interaction is given by Eq. \eqref{eq26} where the interaction is the interaction between opposite spin electrons (which generates the factor of $(-3)$ in the spin dependent term and the weighting factor is the Legendre polynomial $P_0 =1$. As noted earlier, the Fermi surface averaged effective interaction is repulsive for $M/m<1.5$. For the triplet pairing, the paired electrons have the same spin and the factor of $(-3)$ becomes $(+1)$, and the weighting factor is $P_1 = 1 – 0.5 (q/k_F)^2$ \cite{ref26}. This average of the triplet interaction yields the effective interaction for p-wave superconductivity and is shown in Fig. \ref{fig12}.

Figure \ref{fig12} shows that for mass ratios $M/m<1.5$, P-wave superconductivity is allowed, but the net interaction is relatively weak which should result in a low superconducting transition temperature. Note for equal masses, both the electrons and positive fermions would have the same predicted transition temperature. Both species could be superconducting at the same time.

\section{Transport.}

Both the electrons and positive fermions contribute to the transport of electricity and heat. The electrical and thermal resistivities are calculated using coupled Boltzmann equations. Simple kinetic equations using relaxation times for the different scattering mechanisms are adequate for some conditions. See for example Refs. \cite{ref9,ref10,ref11,ref12,ref28,ref29} and further references therein. An electrical resistivity $\rho$ that varies as $T^2$ is characteristic of scattering between degenerate fermions which results from the fact that only fermions with energies within $k_{B}T$ of the Fermi surface are available to conduct electricity and scatter. 

The electron-positive fermion gas considered in this paper is analogous to compensated semimetal. Uncompensated systems are discussed in the references. In the electron-positive fermion gas, the only scattering can be between two electrons, two positive fermions, or an electron with the positive fermion. In a real material, there will be other scattering mechanisms including impurity scattering.

The electrical resistivity is easiest to understand. The electric current carried by the electrons is the electron charge divided by the electron mass times the total momentum of the electron. (the sum of the momentum of the individual electrons). Electron-electron scattering conserves momentum and therefore cannot degrade the electrical current. This is true unless there is another mechanism to conserve momentum such as a reciprocal lattice vector.

The same argument applies to scattering of two positive fermions. This mechanism also does not contribute to the electrical resistivity.

The total current of the electron-positive fermion gas is the sum of the current carried by the electrons and the positive fermions. Because the positive fermions have the opposite charge and a different mass, the total current is not proportional to the total momentum of the system. Therefore electron-positive fermion scattering does contribute to the electrical resistivity. (There is also a contribution to the resistivity if the second species is also negatively charged but with a different mass such as in s-d scattering).

When this system is compensated and the number of electrons equals the number of positive fermions the overall system is neutral and an electric field does not change the total momentum. The field however induces relative momentum between the electrons and positive fermions which results in an electrical current. Electron-positive fermion scattering relaxes the relative momentum to zero and thus contributes to the resistivity. The simple kinetic arguments are presented in Refs. \cite{ref9,ref12}. The temperature dependence becomes more complicated as you pass through the Fermi temperature of the heavier positive fermions. The electrons remain degenerate, but the heavier fermions are in the regime of warm dense matter.

The thermal resistivity is not as easy to understand intuitively. Both the electrons and positive fermions carry heat from the high-temperature side to the low-temperature side. Electron-electron, positive fermion-positive fermion, and electron-positive fermion scattering all contribute to the thermal resistivity. Energy is carried from the hot side to the cold side and there are both inelastic and inelastic collisions. The theoretical approach again involves two coupled Boltzmann equations and the results are expressed in terms of averages of the scattering transition rates. References \cite{ref12,ref17,ref27,ref28,ref29,ref30,ref31} provide details on the Boltzmann equation and screened interactions used to calculate transition rates.

The scattering rates define the relaxation times in the kinetic theory model. The calculation of the scattering rates is an average of the effective potential, and multiple scattering must be taken into account for quantitative calculations. Most previous work used a version of the Thomas Fermi potential with different screening lengths. For electron-electron and positive fermion- positive fermion scattering, the two particles are identical. If they have the same spin, there is a direct and exchange contribution. In most previous papers, the exchange contribution was ignored. In addition, the first Born approximation was generally used. 

The Weidemann-Franz postulates that the ratio of the thermal conductivity to the electrical conductivity is the Lorentz constant times temperature \cite{ref14}.  However the electrical resistivity is only due to electron-positive fermion scattering, while the thermal resistivity in the denominator has contributions from all of the scattering. The ratio should still be proportional to temperature, but the constant of proportionality should be smaller than the Lorentz number. 

The effect of electron-positive fermion scattering on the electrical resistivity depends on the angle of scattering. Forward scattering (small $q$) will only have a small impact on the electric current. The electrons will largely continue in the same direction. However large angle scattering (large $q$) has a much larger effect. For the thermal resistivity, the situation is not as clear because of the role of inelastic scattering in transmitting the energy from the hot side to the cold side.

\begin{figure*}[ht!]
    \centering
    \includegraphics[width=0.48\linewidth]{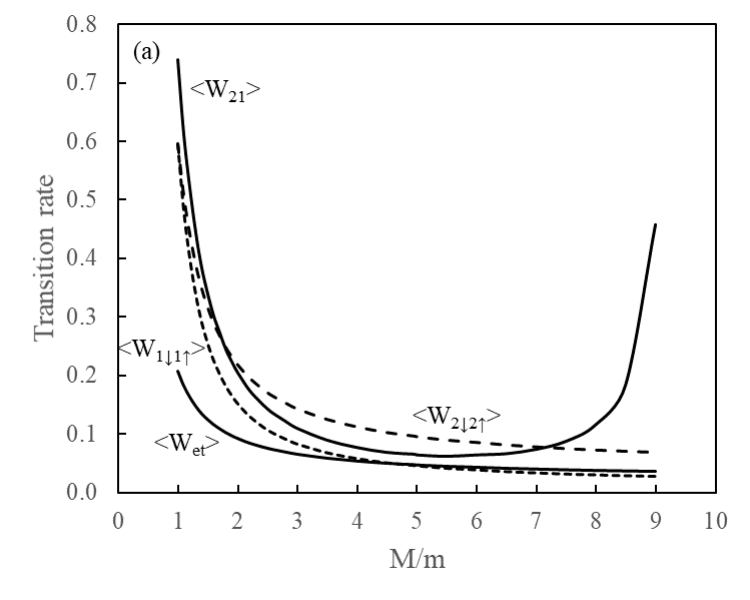}
    \includegraphics[width=0.48\linewidth]{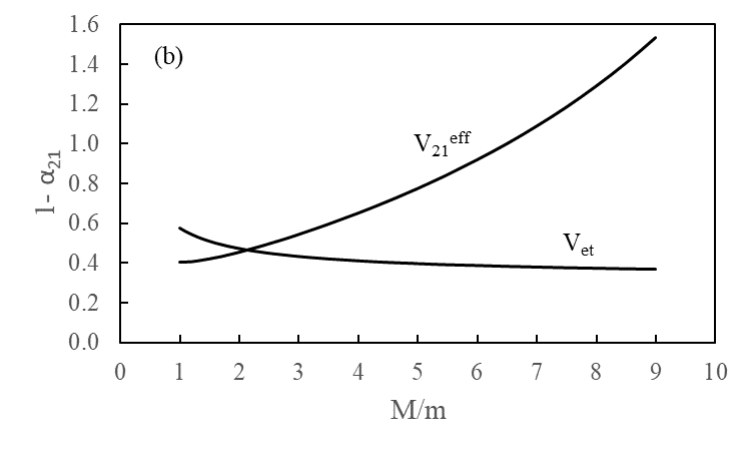}
    \includegraphics[width=0.5\linewidth]{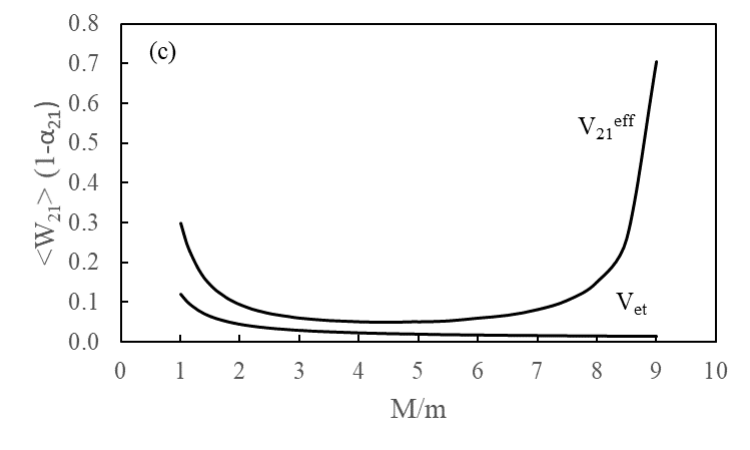}
    \caption{Angular averages of the effective interactions in the electron-positive fermion gas that contribute to the electrical and thermal resistivities. For each mass ratio $M/m$, the calculations are made at the unique density $r_s$ where the energy is a minimum.  Fig. \ref{fig13}a compares the transition (scattering) rates of electron-positive fermion scattering using the effective interaction of this paper to that calculated using the electron-test charge interaction which is close to the Thomas Fermi interaction. Also shown are the transition rates for scattering of opposite spin electrons and opposite spin positive fermions. These do not contribute to the electrical resistivity, but to contribute to the thermal resistivity. Fig. \ref{fig13}b shows the additional angular average factor needed to calculate the electrical resistivity. Fig. \ref{fig13}c is the product of the transition rate and the angular factor. This is proportional to the coefficient of the $T^2$ term in the electrical resistivity. }
    \label{fig13}
\end{figure*}

I do not attempt to make an accurate calculation of the scattering rates or resistivities. My goal is to show the effect of using the many body effective potentials of this paper compared to the results using the electron-test charge interaction (which is similar to the Thomas Fermi approximation). Furthermore, I use the Born approximation which is known to overestimate the scattering cross-section \cite{ref17} by approximately a factor of two. The relevant scattering rates in the Born approximation are given in Appendix B of Ref. \cite{ref12}. For the electron-positive fermion scattering, the coefficient of the $T^2$ term in the electrical resistivity due to the electron-possitive fermion scattering in the compensated system is proportional to certain averages of the transition rate;
\begin{eqnarray}
\rho_{21} &\propto&   \braket{W_{21}} (1- \alpha_{21}) \; ,       
\end{eqnarray}
where the averages of the effective interaction are given by
\begin{eqnarray}
\braket{W_{21}} &=& \frac{1}{2} \int^2_0 dx V^2_{12}(x) \; ,
\end{eqnarray}
and                           
\begin{eqnarray}
\alpha_{21} & = & \frac{1}{2} \int_0^2 dx V^2_{12}(x)\frac{ \left( 1-\frac{x^2}{2} \right)}{ \braket{W_{21}}} \; .
\end{eqnarray}
These quantities are calculated and shown in Fig. \ref{fig13} using both the effective electron-positive fermion interaction given in Eq. \eqref{eq7}, $V_{21}^{\text{eff}} = - v \rho_{1\uparrow}(1- 2G_{12}) / \Delta$, and the electron-test charge interaction $V_{\text{et}} = -v/ \varepsilon_{\text{et}}$. Both interactions are shown in Figs. \ref{fig5} \& \ref{fig6}. Note that the electron-test charge interaction is close to the Thomas Fermi interaction used in previous papers.

Figure \ref{fig13}a compares the angular averages of the electron-positive fermion scattering potential, with that of the potentials for scattering of antiparallel electrons or positive fermions with the electron test charge potential (which is nearly the Thomas Fermi potential) which was used for many previous calculations. Only electron-positive fermion scattering contributes to the electrical resistivity, but all three scattering mechanisms contribute to the thermal resistivity. The angular averages themselves are difficult to interpret without looking back at the potentials themselves in Figs. \ref{fig5} \& \ref{fig6}. At small mass densities, the potentials have a minimum at $q=0$, where at $M/m=4.97$, the influence of the charge density wave at finite $q$ moves the minimum of the potential to wave vectors near and above the Fermi wave vector. Moreover, the density ($r_s$) where the charge density wave first occurs is moving closer to the density where the energy is a minimum ($r_{s \, \text{minimum}}$). These two cross above the mass ratio $M/m=9$, and the appearance of a charge density wave invalidates the basic assumption of the simple model that the background is uniform. This is the reason that the predictions of the simple model terminate at $M/m=9$. Figure \ref{fig13}b shows the additional angular average needed for the electrical resistivity. The large difference between using the electron- positive fermion interaction and the electron test charge interaction is due to the influence of the incipient charge density wave at the larger mass ratios. The overall result for coefficient of the $T^2$ term in the electrical resistivity is shown in Fig. \ref{fig13}c.

\begin{figure}[h!]
    \centering
    \includegraphics[width=1\linewidth]{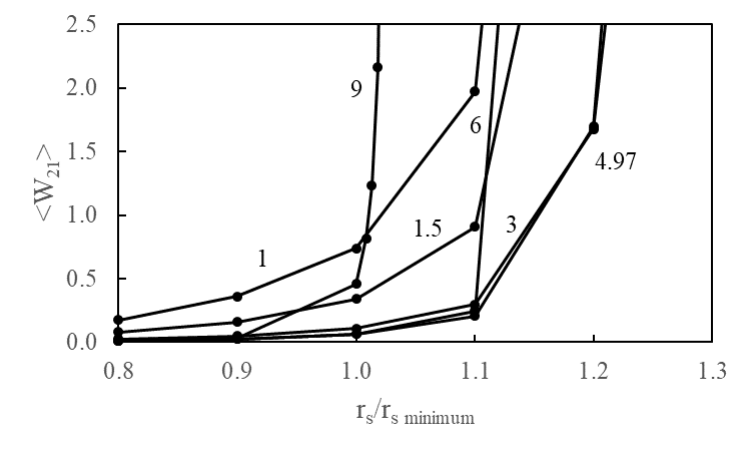}
    \caption{Angular averaged electron-positive fermion transition (scattering) rate of Eq. \eqref{eq26} for the different mass ratios indicated calculated at ratios of the density $r_s$ minimum where the energy is a minimum. }
    \label{fig14}
\end{figure}

Figure \ref{fig13} shows the scattering rates when the electron-positive fermion gas is at the density of the energy minimum. The dependence on $r_s$ is shown is shown in Fig. \ref{fig14} which plots the angular average of the scattering rate at different mass ratios for densities that are between $0.8$ times the $r_s$ of the energy minimum to lower densities (larger $r_s$) near the instability. 

Figure \ref{fig14} reflects the general results of this paper. At high densities that are not near one of the instabilities, the effective interactions (even with proper account of exchange and correlation and the additional screening of the positive fermions) are relatively smooth and featureless. However they are strongly influenced by the incipient compressibility and charge density wave instabilities and diverge at these instabilities.

The steep slope of the $T^2$ resistivity with $r_s$ near an instability has the same behavior as in a calculation of the superconducting transition temperature.

Although there is no mapping of the electron-positive fermion gas onto any real system. This result suggests that a system with a $T^2$ and a high superconducting transition temperature would be very sensitive to pressure which decreases $r_s$ and therefore the resistivity and transition temperature.
 
\section{SUMMARY.}
The electron-positive fermion gas is a simple to state intuitive model of that, in the approximations of this paper, exhibits both $q=0$ (compressibility) and finite q (charge density wave) instabilities. The additional screening from the positive fermions predicts superconductivity and a $T^2$ electrical resistivity. Both are strongly enhanced at densities near the instabilities. 

In the simple model of this paper, the positive fermions are assumed to provide a uniform background for the electrons, and vice versa. With this assumption, the ground state energy of the system is taken to be the energy of the electron gas in a uniform background plus the energy of a positive fermion gas in a uniform background and an unknown additional electron-positive fermion correlation energy. Previous work has shown that this correlation energy is relatively independent of density, so that the pressure and bulk modulus which are derivatives of the energy with respect to volume do not depend on this correlation energy. With this additional assumption, the energy of the system is completely specified within a constant. The uniform electron-positive fermion gas is a solution, but may be metastable or unstable compared to other lower energy options such as a gas of excitons. The goal of this paper is to examine the response functions and the instabilities of the uniform electron-positive fermion gas.

The positive fermion in a uniform background is the same problem as an electron in a uniform background, but with a different mass M. The energy of the electron gas in a uniform background is well known and no new calculations are required for the electron-positive fermion gas. For every mass ratio $M/m$, the system has a unique density $r_s$ where the energy is a minimum. The bulk modulus becomes zero at closely $1.25$ times the $r_s$ of the energy minimum. When the total bulk modulus equals zero, the system is truly unstable.

Linear response theory relates the response functions to unknown local field factors and explicitly includes the coulomb interaction between electrons and positive fermions. A further assumption is that the local field factor is simply that of the electron gas in a uniform background. In principle, there is an additional local field factor due to electron-positive fermion correlation, but that is also ignored in the simple model.

With these assumptions, everything is known and there are explicit formulas for all of the interactions as a function of mass ratio $M/m$ and density $r_s$. All calculations can be done in Excel on a laptop. The problem of two opposite charged fermions was also addressed using density functional theory \cite{ref22} using what amounts to the same assumptions to calculate the phase diagram. The two approaches agree in the region of overlap as was shown in \cite{ref1}.

Ref. \cite{ref13} and references therein, show that the electron gas in a uniform background has no charge or spin instabilities up to $r_s$ greater than $100$. The same is true for positive fermions in a uniform background.

The fact that the positive fermions provide additional screening in addition to screening by the electrons is the crucial new physics element. This additional screening allows superconductivity. More importantly, the local field factor at intermediate wave vectors $1-2 \, k_F$ determined by Quantum Monte Carlo calculations \cite{ref13,ref19} and further references therein show that the local field factor continues to rise as $q^2$ and this combined with the additional screening by the positive fermions gives rise to the charge density waves. This is equivalent to making the local spin density approximation in density functional theory. The quantum Monte Carlo calculations show that the local field factor falls quickly below $q^2$ near $q=2\, k_F$ and the results at these large wave vectors are likely inaccurate.

The simple model predicts the $q=0$ and charge density waves instabilities at finite $q$. It says nothing about the new phases on the other side of the instabilities. The discussion in Ref. \cite{ref22} mentions condensation to a denser state for the $q=0$ instability, and smectic liquid crystals, exotic phases and quantum crystals.

I use the Kukkonen Overhauser approach \cite{ref1,ref3,ref21} to calculate the effective interactions between two electrons, two positive fermions and electrons and positive fermions. These interactions are used to calculate superconductivity and the electrical resistivity of the electron-positive fermion gas.

The simple model predicts that if $r_s$ is near that of a charge density wave, the superconducting transition temperature and the coefficient of the $T^2$ normal state electrical resistivity can become very large.

However there is no mechanism to change $r_s$ without an energy cost. The system is thermodynamically stable or metastable with a minimum energy at a specific $r_s$ for every mass ratio $M/m$.  The simple model uses point positive fermions and coulomb interactions. However the formalism allows for a pseudopotential which could represent core electrons in a real material.

Going beyond the simple model could be difficult. A calculation of the ground state energy including the electron-positive fermion correlation energy would put the $q=0$ instability on firmer grounds.

Quantum Monte Carlo calculations of the response functions of the uniform electron gas are already near the limits of numerical capability and tend not to converge when the density approaches the compressibility instability. The local field factors are extracted from these response functions. The response function and thus the local field factors of the two component Fermi gas may be beyond current capabilities.

Near the phase transitions, the response becomes large and there may be opportunities to go beyond linear response or to consider a renormalization approach.

The simple theory seems consistent for mass ratios $M/m < 9$. At these small ratios the Born-Oppenheimer approximation is questionable.

Most calculations of superconductivity treat the deformable background separately as phonons. In the present case, the positive fermions need to be treated the same way as the electrons for ab initio calculations.

If the mass ratio is modest, the Fermi temperature of the positive fermions is small and the positive fermions could be considered warm dense matter in a background of fully degenerate electrons. This may yield some interesting new physics.

At high temperatures, the positive fermions could be metal ions which would be classical particles immersed in a degenerate electron gas. The ions would still be mobile and provide some additional screening. This might apply to liquid metals.

In Appendix A, I discuss some of the limitations of the simple model and make several speculations. 

A basic prediction from the simple model is that if the mass ratio and density are near an instability, as the electron-positive fermion gas is cooled, the normal state electrical resistivity is large and varies as $T^2$ until the system becomes superconducting at a high temperature. Application of pressure to a system near an instability would strongly reduce the $T^2$ resistivity and the superconducting transition temperature.

\section*{Acknowledgment}
 I thank Daniel Arturo Brito Urbina for preparation of the publication version of this paper.

\appendix
\section{Problems with the simple model, questions and speculation}

The simple model for the electron-positive fermion gas used in this paper allows predictions for all densities and mass ratios. It agrees with results from density functional theory \cite{ref22}, that use essentially the same assumptions.

The fundamental assumptions in the simple model are 
\begin{enumerate}
    \item{The neutralizing background is uniform. }
    \item{The electron-positive fermion correlation energy is constant with respect to volume and therefore does not contribute to the pressure or bulk modulus.}
    \item{The local field factors from the uniform electron gas can be used and any contribution from the electron-positive fermion interactions are ignored.
    \begin{enumerate}
        \item{The local field factor from the high density electron gas at $r_s = 2$ calculated by Quantum Monte Carlo can be scaled to the low density electron gas at $r_s > 20$.} 
    \end{enumerate} }
\end{enumerate}

The simple model predicts the instabilities, but says nothing about what are the new phases on the other side of the instabilities. Are these new phases dramatically different such as a transition to an insulating exciton gas? Or are they less dramatic and can be treated as perturbations to the simple model? Are the new phases compatible with superconductivity? Linear response theory predicts divergences and large responses. A more complete approach considering nonlinear response or renormalization is needed.

The following short discussion summarizes a major point that puzzles me about the simple model.

From thermodynamics, derivatives of the ground state energy with respect to volume yield the pressure which is zero at the equilibrium density ($r_s$), and the bulk modulus which signifies the compressibility instability when it equals zero. There is a unique density $r_s$ of the energy minimum for every mass ratio $M/m$. With only coulomb interactions, the energy minimum ranges from $r_s = 4.18$ for equal masses to $r_s = 1.63$ for an infinite mass positive fermion. The total bulk modulus of the electron-positive fermion gas becomes zero and the system becomes unstable at $r_s$ equal to $1.25$ times the equilibrium $r_s$.

From linear response theory, using the local field factors from the electron gas which have the same $q=0$ instability, the additional screening from the positive fermions predicts a charge density wave that occurs at a different $r_s$ at a certain wave vector for each mass ratio $M/m > 4.97$.

The simple model seems reasonable and consistent until the mass ratio $M/m > 9$. Below that mass ratio, at the $r_s$ of the energy minimum, the gas is stable, and there is no charge density wave. The response functions and effective interactions are influenced by the incipient compressibility and charge density wave instabilities, but the assumption of a uniform system remains valid.

\begin{figure}[h!]
    \centering
    \includegraphics[width=1\linewidth]{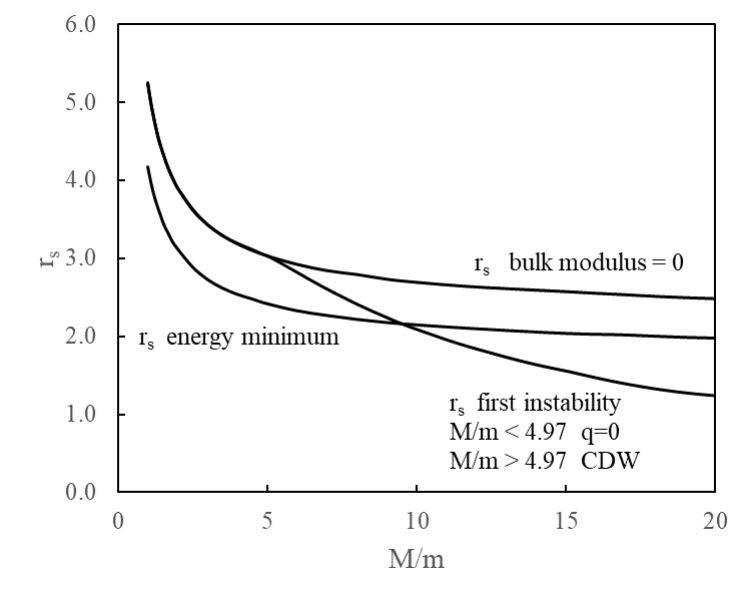}
    \caption{The $r_s$ where the energy of the electron-positive fermion gas is a minimum in the simple model as a function of mass ratio $M/m$, and the $r_s$ where the bulk modulus equals zero and the system becomes unstable. The bulk modulus is calculated by differentiating the energy. The $r_s$ of the instability is closely $1.25$ times that of the energy minimum. Also shown is the $r_s$ where the charge density wave instability first occurs.}
    \label{fig15}
\end{figure}

For $M/m > 9$, the electron-positive fermion gas at the $r_s$ of the energy minimum would always have a charge density wave, and the background is no longer uniform which violates the basic assumption of the simple model. This behavior was reported in Ref. \cite{ref1}, and is shown in Fig. \ref{fig15}.

Is this a real effect? Do the curves actually cross? Can the electron gas local field factors valid at $r_s = 2$ be scaled to large $r_s$?  Does the presence of the charge density wave seriously modify the local field factor at densities near the charge density wave and particularly at wave vectors near $2\,k_F$? Or does the charge density wave seriously impact the energy and the calculated equilibrium $r_s$? These are some unanswered questions.

For $M/m=9$, the equilibrium density $r_s = 2.18$. However the effective density of the positive fermions is larger by the factor $M/m$ and equals $19.6$. The simple model assumes that the local field factor at high density can be scaled to low density. There are no quantum Monte Carlo data at this low density to verify this assumption.

It seems to me that a small amplitude charge density wave should not change the ground state energy substantially. That means it would have little effect on the equilibrium density and compressibility instability, and I speculate that the system could be stable with an $r_s$ close to the predicted energy minimum and also have a charge density wave. Superconductivity can coexist with charge density waves. See \cite{ref34,ref35,ref36} for example.

Another assumption of the simple model is that the electron-positive fermion correlation contribution to the local field factor $G_{12}$ can be ignored. The $q^2$ dependence of the local field factor near $q = 1-2 \, k_F$ is the crucial to the existence of the charge density wave. The Hubbard model for the local field factor does not have charge density waves at all. The $q^2$ dependence at intermediate $q$ is predicted by Quantum Monte Carlo calculations. The local field factor is a result of a local approximation to the many body problem and I do not have a feeling for the underlying microscopic physics. It could be that the missing correlation contribution, or the charge density wave strongly affects the local field factor at intermediate $q$.

Can the simple model be mapped onto real materials?  Or can the predictions of the simple models elucidate the behavior of charge density systems?

A prediction of the simple model is that near the charge density wave, the system can be very sensitive to $r_s$. This would imply sensitivity to pressure. A system with a charge density wave should find that the charge density wave vector increases with pressure and finally charge density wave disappears entirely. This behavior is illustrated in Fig. 16 that is taken from Ref. \cite{ref1}.

\begin{figure}[h!]
    \centering
    \includegraphics[width=1\linewidth]{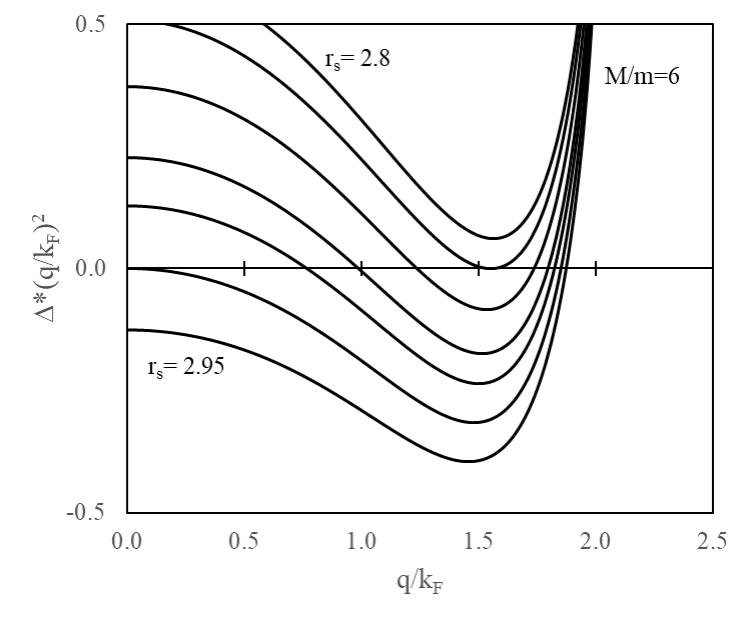}
    \caption{Plot of the denominator of the induced density as a function of $q/k_F$ for $M/m = 6$. The second curve from the top is the onset of the charge density wave at $r_s = 2.821$. The third curve from the top is at $r_s = 2.85$. The wave vector of the charge density wave is where the curve crosses zero. }
    \label{fig16}
\end{figure}

The discussion in Section VIII, also shows that near the instability, the predicted $T^2$ electrical resistivity and the superconducting transition temperature are very sensitive to $r_s$ which implies sensitivity to pressure.

I speculate that the compressibility instability at $q=0$ could be a metal-insulator transition. This instability occurs as $r_s$ increases or density decreases. In the opposite case, an insulator near the metal insulator transition could become a metal under pressure which increases density. An interesting experimental fact is that silicon, a semiconductor, becomes a metal upon melting \cite{ref35,ref36}. In addition, it’s density increases ($r_s$ decreases) upon melting, just as when ice melts. This is why ice cubes float on water. A wild speculation is that this change in density is what triggers the semiconductor-metal transition. In the simple model, the silicon ions would be classical particles and a pseudo-potential would be needed.

Can the results of the simple model for a uniform background be used somehow as the basis for calculations for nonuniform systems? This would be in analogy to using the uniform electron gas results for calculations of nonuniform systems in density functional theory. For example, at a surface, edge or some internal structure, the density may vary between a high density and zero or a low density, and pass through the critical density for localized instabilities. Of course, this may require the straightforward extension of the simple model to two or one dimension. A localized proximity to a charge density wave might also enhance the possibility of superconductivity in that region.

\newpage
\bibliography{reference}
\end{document}